\documentclass[a4paper,11pt]{article}

\pdfoutput=1 

\usepackage{jheppub, amsmath, amssymb,amsthm,dsfont,graphicx,mathtools,multirow,rotating,placeins,verbatim}

\usepackage{mathrsfs} 

\usepackage{xcolor}
\usepackage{slashed}
\usepackage{tensor}
\usepackage[T1]{fontenc} 
\usepackage{float}
\usepackage{tikz-cd} 
\usepackage{tikz} 
\usepackage{todonotes} 
\usepackage{cases} 
\usepackage[utf8]{inputenc} 
\usepackage{hyperref} 
\usepackage{cleveref} 
\usepackage{mathrsfs} 
\usepackage{cancel} 
\usepackage[percent]{overpic} 
\usepackage{longtable} 
\usepackage{array} 
\usepackage{booktabs} 
\usepackage{tabularx} 
\usepackage{colortbl} 
\usepackage{xcolor} 
\usepackage{hhline} 
\usepackage{float} 
\usepackage{amsmath} 
\usepackage{amsthm} 
\usepackage{empheq} 
\usepackage{bbm} 
\usepackage{amsfonts} 
\usepackage{calligra} 
\usepackage{braket} 
\usepackage{enumitem} 
\usepackage[english]{babel} 
\usepackage{xcolor} 
\usepackage{stmaryrd} 
\usepackage{csquotes} 
\usepackage{subcaption} 
\usepackage{MnSymbol} 
\usepackage{diagbox} 
\usepackage{makecell} 
\usepackage{xspace} 
\usepackage{tabularx} 
\usepackage{colortbl}

\crefname{subsubsection}{section}{sections} 
\crefname{lemma}{lemma}{lemmas} 

\newcommand{\be}{\begin{equation}}
\newcommand{\ee}{\end{equation}}

\DeclareMathAlphabet\mathbfcal{OMS}{cmsy}{b}{n}

\setuptodonotes{linecolor=black!50,backgroundcolor=yellow!80,size=footnotesize}

\usepackage{leftindex}

\title{Carroll Double Copy}

\author[a]{Sucheta Majumdar,}
\author[b]{Silvia Nagy,}
\author[b]{Giorgio Pizzolo}
\affiliation[a]{Department of Physics and Electronics, Christ University, \\Hosur Road, Bangalore 560029, India}

\affiliation[b]{Department of Mathematical Sciences, Durham University, Durham, DH1 3LE, UK}

\emailAdd{giorgio.pizzolo@durham.ac.uk, silvia.nagy@durham.ac.uk, sucheta.majumdar@christuniversity.in }

\abstract{We construct a double copy for Carroll gravity as the ``square'' of Carroll Yang–Mills theory at the linearised level. Starting from the convolutional double copy prescription in Lorentzian spacetime, we take the Carrollian limit $c\to 0$ to establish a double copy dictionary compatible both with symmetries and the corresponding equations of motion. Since this limit yields two distinct sectors, namely the electric and magnetic Carroll theories, we systematically investigate double copy constructions involving all possible combinations of these parent theories. To accommodate the time-independent on-shell configurations of the magnetic Carroll fields, we introduce a suitable regularisation procedure for the convolutional double copy. In addition, the covariance of the double copy under Carroll boosts requires an appropriate dressing of the fields, leading to the inclusion of a Goldstone-type object in the construction. We first develop the formalism using toy models involving scalar fields and then apply it to the double copy between linearised Carroll Yang–Mills theory and Carroll gravity.}
\begin{document} 

\maketitle


\newpage
\section{Introduction}

Carrollian field theories, obtained as the $c\to0$ limit of relativistic field theories, have been attracting increasing interest in the context of flat space holography. This is due to the Carrollian geometry of null infinity, where the Bondi–Metzner–Sachs (BMS) symmetry group admits an interpretation as a conformal Carroll group~\cite{Duval:2014uva}. Carroll symmetry also appears in a wide range of other settings, from string theory and ultrarelativistic hydrodynamics to gravitational waves and condensed matter systems.\footnote{For some recent reviews see~\cite{Bergshoeff:2022eog,Ruzziconi:2026bix,Bagchi:2025vri} and references therein.}  

The study of scattering amplitudes provides a concrete realisation of the holographic correspondence, via their mapping to boundary correlation functions~\cite{Bagchi:2022emh,Alday:2024yyj}. In this context, another type of duality, namely the double copy correspondence and associated Bern-Carrasco-Johansson (BCJ) relations~\cite{Bern:2008qj,Bern:2010ue,Bern:2019prr}, has helped to shed light on the structure of gravity amplitudes via their unexpected connection to gauge theory amplitudes. Recent work has started to reveal double copy and BCJ structures in the context of Carrollian amplitudes~\cite{Surubaru:2025fmg,Bagchi:2026pjx,Chakrabortty:2024bvm,Stieberger:2024shv}. 

A natural question is if the double copy is present at a fundamental level, i.e.\ directly at the level of the fields, their equations of motion and symmetries (both local and global). This can be answered in the positive, in the linearised theory, where we can explicitly realise the double copy via a convolution product~\cite{Anastasiou:2018rdx,Anastasiou:2014qba,Godazgar:2022gfw,Beneke:2021ilf,Ferrero:2020vww,Borsten:2020xbt,Luna:2020adi,LopesCardoso:2018xes,Cardoso:2016amd,Cardoso:2016ngt}. Crucially, this allows for natural generalisations beyond Lorentzian flat-space, namely 
homogeneous spaces~\cite{Borsten:2019prq,Borsten:2021zir}, (A)dS~\cite{Liang:2023zxo} and at null infinity to study supertranslations as double copy of large gauge symmetries~\cite{Ferrero:2024eva}. For other forms of the double copy beyond flat Lorentzian backgrounds, see references in~\cite{CarrilloGonzalez:2026phk}\footnote{We note that an interesting connection between Carroll symmetry and the classical double copy also arises for gravitational plane waves \cite{Duval:2017els,Ilderton:2018lsf}.}. 

In this article, we take the first few steps towards establishing the convolutional double copy in the Carrollian limit. Our starting point is the double copy dictionary on a Lorentzian background, defined as a double convolution
\be\label{convo_intro} 
h\coloneqq f^a\ast(\varphi^{-1})^{a\tilde b}\ast g^{\tilde b} \,.
\ee 
In the above, $(\varphi^{-1})^{a \tilde b}$, denoted as the spectator, is the convolutional inverse of a biadjoint scalar and $f^a$ and $g^{\tilde b}$ are fields of helicity $\text{h}_f$ and $\text{h}_g$ valued in the non-abelian groups $G$ and $\tilde G$, respectively. The outcome will be a field $h$ of helicity $\text{h}_g+\text{h}_f$. This, of course, covers the construction of the graviton from two gauge fields, but it also allows for other combinations (for example a scalar is the double copy of two scalars, and a gauge field is obtained from a scalar and another gauge field) which we will employ as toy models in order to set up our dictionary. 

We develop the Carrollian limit of \eqref{convo_intro}, which leads to a double copy dictionary compatible with the Carrollian global symmetries, local symmetries, as well as the equations of motion (E.o.M.). This limit can be consistently taken in two possible ways, leading to the so-called electric and magnetic branches. These are arrived at by working in the Hamiltonian formalism and rescaling the  field and its conjugate momentum as $(\phi,\pi)\to(c\phi,c^{-1}\pi)$ in the electric case and $(\phi,\pi)\to(\phi,\pi)$ in the magnetic case, and then taking the $c\to0$ limit.   We find that the choice of Carrollian branch on the input fields $f$ and $g$, as well as the spectator $\varphi$ determines the Carrollian branch of the double copy field $h$ as
\begin{equation*}
    \frac{\text{E}_f\times\text{E}_g}{\text{E}_\varphi},\ \frac{\text{E}_f\times\text{M}_g}{\text{M}_\varphi}\ \to\ \text{E}_h\;, \quad \frac{\text{E}_f\times\text{M}_g}{\text{E}_\varphi},\ \frac{\text{M}_f\times\text{M}_g}{\text{M}_\varphi}\ \to\ \text{M}_h\;, \quad \frac{\text{E}_f\times\text{E}_g}{\text{M}_\varphi},\ \frac{\text{M}_f\times\text{M}_g}{\text{E}_\varphi}\ \to\ \varnothing\;,
\end{equation*}
where we do not list the cases obtained by symmetry of the convolution. We see that mixed choices are allowed and can lead to well-defined limits. Gratifyingly, the final two combinations which are forbidden by the $c$ rescalings, turn out to also be ill-defined mathematically.  

As a by-product of our dictionary, we also extend the remit of the convolutional double copy in a mathematically rigorous way. In particular, we show how to regularise the convolution product in cases where the inputs are time independent. This confirms previous double copy constructions for classical solutions~\cite{Luna_2020}, while establishing them on a more solid mathematical foundation.  

Finally, in order to preserve the Carroll symmetries, particularly the boosts, via the double copy, we find that for some of the cases above we need to introduce a (constant) \emph{dressing field} in the convolution. These dressing fields, which behave as Goldstone (or Stueckelberg) fields, are becoming ubiquitous in the study of asymptotic symmetries~\cite{Cresto:2024mne,Cresto:2024fhd,Javadinezhad:2022hhl,Agrawal:2025bsy,Nagy:2025hip,Nagy:2024jua,Nagy:2024dme,Nagy:2022xxs,Campiglia:2021oqz}. In our case, the dressing field does not introduce additional degrees of freedom, consistent with the fact that the Carrollian group is an Inonu-Wigner contraction of the Poincar\'e group~\cite{Bacry:1968zf}, and thus has the same dimension. 

The article is organised as follows. In Section \ref{sec:EM_limits}, we present a short review of the Carroll limit of Lorentzian field theories, focusing on massless scalar fields, linearised Yang–Mills theory and linearised gravity. In Section \ref{sec:Carroll double copy with scalars}, we develop our Carroll double copy construction using some toy models, where we highlight the subtleties associated with defining convolutions on Carroll manifolds. In Section \ref{sec:Carrollian double copy to linearised gravity}, we apply this framework to construct the double copy between linearised Carroll Yang–Mills theory and linearised Carroll gravity. We conclude in \ref{sec:Conclusions} with some remarks on our results as well as outline some future directions.

\section{Carroll limits with sources}
\label{sec:EM_limits}
This section briefly reviews the procedure for obtaining Carroll actions from their Lorentzian counterparts in the $c \to 0$ limit~\cite{Henneaux:2021yzg,deBoer:2021jej,Hansen:2021fxi}. Working within the Hamiltonian formulation, we illustrate the method for scalar fields, Maxwell theory, and linearised gravity, including their couplings to external sources. Since our goal is to construct double copies with these theories, the inclusion of external sources is essential for defining well-behaved convolutions involving the resulting Carrollian fields~\cite{LopesCardoso:2018xes}.
\par
We work in four-dimensional Minkowski spacetime with Cartesian coordinates $(x^0,x^i)$ and introduce the time coordinate $t$ through $x^0 = ct$. In coordinates $(t,x^i)$, the metric takes the form
\begin{equation}
ds^2 = -c^2 dt^2 + \delta_{ij}dx^i dx^j\,.
\end{equation}
In the $c \to 0$ limit, the Carroll geometry is characterized by a degenerate metric $g = \operatorname{diag}(0,1,1,1)$ and a vector $\mathbf{n} = \partial_t$ which lies in its kernel. The isometries preserving the Carroll structure are given by
\begin{equation}\label{Carroll_transformations}
\left(x^\mu\right)'=x^\mu-\xi^\mu,\qquad
    \xi^0  = a^0+ b_ix^i\,,\quad \xi^i = a^i + \omega^i_j x^j \,, 
\end{equation}
such that
\begin{equation}
    \mathcal L_\xi g = 0\,, \quad \mathcal L_\xi \mathbf n = 0\,.
\end{equation}
Here, $a^0, a^i$ stand for the time and spatial translation parameters; $\omega_{ij}$ parameterises the rotations and $b_i$ the Carroll boosts.
\par
In order to derive the Carrollian actions in the $c\to 0$ limit, we adopt the canonical formulation, in which the symplectic structure of phase space provides a natural and systematic framework for taking this singular limit.
\subsection{Scalar theory}

  We first consider the Lorentzian action for a massless scalar in four dimensions. Upon restoring the factors of $c$, the massless Klein-Gordon action takes the form

\begin{equation}
    \label{eq:scalar_lagrangian}
    S=\int c\,dtd^3x\,\mathcal{L}=\int c\,dtd^3x\left(\frac{1}{2c^2}(\partial_0\phi^\mathrm{L})^2-\frac{1}{2}(\partial_i\phi^\mathrm{L})^2-j^\mathrm{L}\phi^\mathrm{L}\right)\;,
\end{equation}
where we used the superscript $\mathrm{L}$ to denote Lorentzian quantities, in order to differentiate them from their counterparts after taking the Carrollian limits. The equation of motion is $\Box\phi^\mathrm{L}=j^\mathrm{L}$, where $\Box=\partial_\mu\partial^\mu=-\frac{1}{c^2}\partial_0^2+\partial_i^2$. The conjugate momentum is defined as
\begin{equation}
    \pi^\mathrm{L}=\frac{\delta\mathcal{L}}{\delta(\partial_0\phi^\mathrm{L})}=\frac{1}{c^2}\dot\phi^\mathrm{L}\;,
\end{equation}
so the canonical Hamiltonian density reads
\begin{equation}
    \mathcal{H}=\pi^\mathrm{L}\dot\phi^\mathrm{L}-\mathcal{L}=\frac{c^2}{2}\left(\pi^\mathrm{L}\right)^2+\frac{1}{2}(\partial_i\phi^\mathrm{L})^2+j^\mathrm{L}\phi^\mathrm{L}\;.
\end{equation}

The action in Hamiltonian form is then
\begin{equation}
    \begin{aligned}
        \label{eq:scalar_Hamiltonian}
        S_{H}&=\int c\,dtd^3x\,(\pi^\mathrm{L}\dot\phi^\mathrm{L}-\mathcal{H}) \,.
    \end{aligned}
\end{equation}
 In the $c\to 0$ limit, the Carrollian action is obtained  as the leading order term in the Lorentzian action $S_H = c \,S_\mathrm{Carr} $ with
\begin{equation} \label{eq:Carroll_scalar_hamiltonian_action}
    S_\mathrm{Carr}=\int d^4x\left(\pi\dot\phi-\mathcal{H}_\mathrm{Carr}\right)\,.
\end{equation}
\par 
In the above, $\phi$ and $\pi$ denote the $c\to 0$ limits of $\phi^\mathrm{L}$ and $\pi^\mathrm{L}$. We take this limit in the Hamiltonian phase space in a way that preserves the symplectic structure 
\begin{equation}
    \Omega_{scalar} = \int d^3x \, \delta\pi^\mathrm{L} \wedge \delta \phi^\mathrm{L} \,.
\end{equation}

There are two choices that preserve the symplectic form and lead to finite actions, giving two kinds of Carrollian theories commonly referred to as the magnetic and electric branches~\cite{Henneaux:2021yzg,deBoer:2021jej}. We briefly describe these two limits below:
\begin{itemize}
    \item \textit{Magnetic case:} In this case, the Lorentzian fields are mapped to the Carrollian fields without any factor of $c$
\begin{equation}\label{E_sc}
    \phi^\mathrm{L}=\phi\;, \quad \pi^\mathrm{L}=\pi\;, \quad j^\mathrm{L}=j \,.
\end{equation}
In the $c \to 0$ limit, the magnetic Hamiltonian density is given by
\begin{equation}
    \label{H_Mag}\mathcal{H}^\mathrm{M}_\mathrm{Carr} = \frac12(\partial_i\phi)^2+j\phi\,,
\end{equation}
where the $c$ scaling of $j$ is justified a posteriori by the fact that we want to keep the source in the Hamiltonian after taking the Carrollian limit. The action becomes
\begin{equation}
        S^\mathrm{M}_\mathrm{Carr}=\int \,d^4x\left[\pi\dot\phi-\frac{1}{2}(\partial_i\phi)^2-j\phi\right] \;.
    \end{equation}
  with the equations of motion
\be 
\label{magEOM}
\dot{\phi} = 0\,,\quad \dot{\pi} = \partial_i\partial^i\phi -j \,.
\ee
    \item \textit{Electric case:} In the electric case, the Lorentzian field $\phi$ and momentum $\pi$ are rescaled inversely as follows
\begin{equation}
    \phi^\mathrm{L}=c\phi\;, \quad \pi^\mathrm{L}=\frac{\pi}{c}\;, \quad j^\mathrm{L}=\frac{j}{c} \,.
\end{equation}
 such that in the $c \to 0$ limit, the electric Hamiltonian density reads 
\begin{equation}
    \label{H_Elec}\mathcal{H}^\mathrm{E}_\mathrm{Carr} = \frac12\pi^2+j\phi\,.
\end{equation}
with action
\begin{equation}
        S^\mathrm{E}_\mathrm{Carr}=\int \,d^4x\left[\pi\dot\phi-\frac{1}{2}\pi^2-j\phi\right] \;.
    \end{equation}
    with the equations of motion
\be
 \dot{\phi} = \pi \,,\quad \dot{\pi} = -j \,.\label{elecEOM} 
\ee
It is interesting to note that in the electric case the symplectic structure only partially constrains the rescalings. One could rescale $\phi^\mathrm{L}$ by $c^n$ for any $n$, provided the momentum $\pi^\mathrm{L}$ is rescaled inversely as $c^{-n}$. However, requiring the resulting Hamiltonian action to be finite imposes an additional constraint that only for $n=0,1$ is the limit non-divergent.
\end{itemize}
A key distinction between electric and magnetic Carroll dynamics lies in their time dependence. In the magnetic case, the fields $\phi$ are time-independent or static, while time dependence enters only through their conjugate momenta $\pi$, as evident from the equations of motion \eqref{magEOM}. As a result, the momenta are not determined by the velocities $\dot{\phi}$ and cannot be eliminated using their own equations of motion. Electric Carroll dynamics, by contrast, more closely resembles the Lorentzian case, in which the conjugate momenta are related to the velocities. This distinction will play a crucial role in defining well-behaved convolutions on the Carrollian manifold that accommodate both time-dependent and static fields.
\par 
Furthermore, the magnetic and electric Carroll fields also differ in their transformation rules under the Carroll boosts given in \eqref{Carroll_transformations}. The electric field and its conjugate momentum transform as
\be \label{el_boost_scalar}
\delta_b \phi= b_i x^i\, \dot{\phi}, \qquad
\delta_b \pi=b_i x^i\, \dot{\pi} \ ,
\ee 
whereas in the magnetic branch, we have
\be 
\label{eq:Carroll_boost_magnetic_momentum}
\delta_b \phi= b_i x^i\, \dot{\phi}, \qquad
\delta_b \pi=b_i x^i\, \dot{\pi} +b^i\partial_i\phi \ .
\ee 
We also note the finite boost transformation for the scalar for later use
  \begin{equation}
     \label{eq:scalar_carroll_boost}
     (\Lambda_b\phi)(t,x^i)=\phi(t+b_jx^j,x^i)\;.
 \end{equation}





\subsection{Linearised Yang-Mills theory}
In this section, we discuss the Carroll limits of the Lorentzian Yang-Mills action with the non-Abelian field strength
\begin{equation}
    F_{\mu \nu}^{\mathrm{L},a} = \partial_\mu A^{\mathrm{L},a}_\nu - \partial_\nu A^{\mathrm{L},a}_\mu + g f^{abc} A^{\mathrm{L},b}_\mu A^{\mathrm{L},c}_\nu\,,\quad A^{\mathrm{L}}_\mu = A^{\mathrm{L},a}_\mu T^a\,,
\end{equation}
where the gauge indices $a,b,c$ run from $1,2,\ldots, N=\text{dim}(G)$, with G the gauge group. As before, the superscript L has been used for Lorentzian fields. Since we are interested in the double copies leading to linearised gravity, we will restrict the discussion to Yang-Mills theory at the linearised level. To that end, we truncate the field strength to linear order
\begin{equation}
     F_{\mu \nu}^{\mathrm{L},a} = \partial_\mu A^{\mathrm{L},a}_\nu - \partial_\nu A^{\mathrm{L},a}_\mu\,.
\end{equation}
The gauge field is treated as a set of free fields and the non-Abelian self-interactions are dropped. The theory comprises $N$ copies of Abelian Maxwell theories, now transforming only globally under the group G:
\begin{equation}
   \mathcal S_{\mathrm{linYM}} = -\frac{1}{4}\int d^4x \,F_{\mu\nu}^{\mathrm{L}} F^{\mathrm{L},\mu\nu } \,.
\end{equation}
In what follows, we shall suppress the gauge index for convenience, unless otherwise specified. The gauge symmetry becomes Abelian in this case 
\begin{equation}
 \label{Gtransfo}
     \delta_{\lambda}A^\mathrm{L}_{\mu} = \partial_\mu \lambda\,.
\end{equation}
We couple a source term to the Lorentzian action 
\begin{equation}
    \mathcal{S}_{\mathrm{linYM}}= -\int d^4x\left(\frac{1}{4}F^{\mathrm{L},\mu\nu}F^{\mathrm{L}}_{\mu\nu}+j^{\mathrm{L},\mu} A^{\mathrm{L}}_\mu \right)\,.
\end{equation}
Restoring the factors of $c$ in the above, we find
\begin{equation}
 \label{eq:linYM}
    \mathcal{S}_{\mathrm{linYM}}= \int c\,dt\, d^3x\left(\frac{1}{2c^2}(F^{\mathrm{L}}_{0i})^2-\frac{1}{4}(F^{\mathrm{L}}_{ij})^2+\frac{1}{c^2}j^{\mathrm{L}}_0A^{\mathrm{L}}_0-j^{\mathrm{L}}_iA^{\mathrm{L}}_i\right) \,.
\end{equation}
The conjugate momentum 
\begin{equation}
\pi^{\mathrm{L},\mu} \coloneqq \frac{\delta\mathcal{L}}{\delta\partial_0A^{\mathrm{L}}_\mu}\;,
\end{equation}
leads to
\begin{equation}
    \pi^{\mathrm{L}}_i=\frac{1}{c^2}F^{\mathrm{L}}_{0i}\;,
\end{equation}
and the primary constraint $\pi^\mathrm{L}_0 = 0$. From the requirement $\dot\pi^\mathrm{L}_0 =0$, we obtain a first-class constraint    $ \mathcal G = \partial_i \pi^\mathrm{L}_i+c^{-2}j_0^\mathrm{L} \approx 0$, which is associated with the gauge freedom in the theory. The canonical Hamiltonian density is
\begin{equation}
\mathcal{H}
  ~=~\frac{c^2}{2}(\pi^{\mathrm{L}}_i)^2+\frac{1}{4}(F^{\mathrm{L}}_{ij})^2-A^{\mathrm{L}}_0\left(\partial_i\pi^{\mathrm{L}}_i+\frac{1}{c^2}j^{\mathrm{L}}_0\right)+j^{\mathrm{L}}_iA^{\mathrm{L}}_i \,.
\end{equation}
The action in Hamiltonian form is
\begin{equation}
\label{YM_SHam}
        S_{H}=\int c\,d^4x\left[\pi^{\mathrm{L}}_i\dot A^{\mathrm{L}}_i-\left(\frac{c^2}{2}(\pi^{\mathrm{L}}_i)^2+\frac{1}{4}(F^{\mathrm{L}}_{ij})^2-A^{\mathrm{L}}_0\left(\partial_i\pi^{\mathrm{L}}_i+\frac{1}{c^2}j^{\mathrm{L}}_0\right)+j^{\mathrm{L}}_iA^{\mathrm{L}}_i\right)\right]\,.
\end{equation}
The canonical gauge generator is given by
\begin{equation}
    Q_\lambda\coloneqq\int d^3x\,\mathcal{G}(x)\lambda(x) \,,
\end{equation}
which, through the Poisson brackets
\begin{equation}
  \{A^\mathrm{L}_{i}(x^k),\pi^\mathrm{L}_{j}(y^k)\}=\delta_{ij}\delta^{(3)}(x^k-y^k)\;, 
\end{equation}
generates the gauge transformations \eqref{Gtransfo} 
\begin{equation} 
    \delta_\lambda A^{\mathrm{L}}_i=\{Q_\lambda,A^{\mathrm{L}}_i\}=\partial_i\lambda\;, \quad \delta_\lambda \pi^{\mathrm{L}}_i=\{Q_\lambda,\pi^\mathrm{L}_i\}=0\,,
\end{equation}
that remain the same in the two Carroll branches. The transformation of the Lagrange multiplier $A_0$ is obtained from the gauge invariance of the Hamiltonian action.

We now take the $c\to0$ limit of the linearised Yang-Mills action~\eqref{YM_SHam} in a way that preserves the symplectic form of the theory
\begin{equation}
    \Omega_{\mathrm{linYM}} = \int d^3x \,\delta \pi^{\mathrm{L}}_i \wedge \delta A^{\mathrm{L}}_i \,.
\end{equation}
The rescaling for the Lagrange multiplier $A_0$ is chosen so as to keep the gauge constraint $\mathcal G$ term in the resulting action. We also choose the rescaling of the external source such that all components of $j^\mu$ survive in the limit~\cite{patil2025,Manfredi_2013}. The two Carroll limits so obtained are described below.
\begin{itemize}
    \item \textit{Magnetic case:} Here, we map the Lorentzian fields to the Carrollian ones as follows
    \begin{equation}
        A^{\mathrm{L}}_0 =  A_0\;, \quad A^{\mathrm{L}}_i =  A_i\;, \quad \pi^{\mathrm{L}}_i = \pi_i\;, 
    \end{equation}
    and the sources as
    \begin{equation}
        j^{\mathrm{L}}_0 = c^2j_0\;, \quad j^{\mathrm{L}}_i = j_i \;.
    \end{equation}
In the $c\to 0$ limit, the Lorentzian action reduces to the magnetic Carroll action
    \begin{equation}
        S^{\mathrm{M}}_{\mathrm{Carr}}=\int \,d^4x\left[\pi_i\dot A_i-\frac{1}{4}(F_{ij})^2+A_0\left(\partial_i\pi_i+j_0\right)-j_iA_i\right] \;.
    \end{equation}
  The equations of motion that follow from this action are
    \begin{equation}
        \label{eq:eom Maxwell magnetic limit}
        \begin{aligned}
            \dot A_i-\partial_i A_0&=0\,, \\
            -\dot\pi_i+\partial_jF_{ji}&=j_i \,, \\
            -\partial_i\pi_i&=j_0\,.
        \end{aligned}
    \end{equation}
    In the source-free case, they are identical to the ones obtained in~\cite{Henneaux:2021yzg}. 
    \item \textit{Electric case:} To obtain the electric sector, we rescale the Lorentzian fields and conjugate momenta inversely as
    \begin{equation}
        A^{\mathrm{L}}_0 =  cA_0\;,\quad A^{\mathrm{L}}_i =  cA_i\;, \quad \pi^{\mathrm{L}}_i = \frac{\pi_i}{c}\;,
    \end{equation}
    and the sources as
    \begin{equation}
       j^{\mathrm{L}}_0 = cj_0\;, \quad j^{\mathrm{L}}_i = \frac{j_i}{c} \;.
    \end{equation}
    The electric Carroll action reads
    \begin{equation}
        S^\mathrm{E}_\mathrm{Carr}=\int \,d^4x\left[\pi_i\dot A_i-\frac{1}{2}(\pi_i)^2+A_0\left(\partial_i\pi_i+j_0\right)-j_iA_i\right] \;.
    \end{equation}
    with the equations of motion
    \begin{equation}
        \begin{aligned}
           \dot A_i-\partial_iA_0&= \pi_i \,,\\
            -\dot\pi_i&=j_i\,, \\
            -\partial_i\pi_i&=j_0 \,.
        \end{aligned}
    \end{equation}
    which are also consistent with the ones obtained in~\cite{Henneaux:2021yzg} for the source-free case.
    \end{itemize}
In linearised Yang-Mills theory as well, the electric and magnetic Carroll
fields transform differently under Carroll boosts. In the magnetic case, the fields transform as 
\begin{equation}
\delta_b A_i=b_jx^j\,\dot A_i+b_iA_0 \,,\quad  \delta_b \pi_i = b_jx^j\,\dot\pi_i+b^jF_{ji}\,,
\end{equation}
whereas in the electric case, the transformation laws are
\begin{equation}
    \label{eq:boost_electric_YM}
 \delta_b A_i= b_jx^j\,\dot A_i+b_iA_0\,, \quad \delta_b \pi_i =  b_jx^j\,\dot\pi_i \,.
\end{equation}
In both cases the Lagrange multiplier $A_0$ transforms as $\delta_b A_0 = b_j x^j\,\dot A_0$.

\subsection{Linearised gravity}

In this section, we describe the procedure to obtain Carroll actions for linearised gravity from the Lorentzian Pauli-Fierz action. We consider the Pauli-Fierz action for a spin-2 field, obtained by perturbing the Einstein-Hilbert action around the flat Minkowski metric $g_{\mu\nu}=\eta_{\mu\nu}+h^\mathrm{L}_{\mu\nu}$. We couple the action to a source $T^\mathrm{L}_{\mu\nu}$ which is symmetric and conserved, i.e.\  $T^\mathrm{L}_{\mu\nu}=T^\mathrm{L}_{\nu\mu}$ and $\partial_\mu T^{\mathrm{L},\mu\nu}=0$. The Pauli-Fierz action with the factors of $c$ restored reads
\begin{equation}
    S^\mathrm{L}_\mathrm{PF}=\frac{c^3}{16\pi G}\int c\,dt \,d^3x\left(-\frac{1}{4}\partial_\rho h^\mathrm{L}_{\mu\nu}\partial^\rho\bar h^{\mathrm{L},\mu\nu}+\frac{1}{2}\partial_\mu\bar h^{\mathrm{L},\mu\nu}\partial^\rho\bar h^\mathrm{L}_{\rho\nu}+\frac{8\pi G}{c^4}h^\mathrm{L}_{\mu\nu}T^{\mathrm{L},\mu\nu}\right) \,,
\end{equation}
where the spacetime indices are raised with $\eta^{\mu\nu}$ and
\begin{equation}
    \partial^0=-\frac{1}{c^2}\partial_t\;, \quad h^\mathrm{L}=-\frac{h^\mathrm{L}_{00}}{c^2}+h^\mathrm{L}_{ii}\;, \quad \bar h^\mathrm{L}_{\mu\nu}=h^\mathrm{L}_{\mu\nu}-\frac{1}{2}\eta_{\mu\nu}h^\mathrm{L} \,,
\end{equation}
where repeated indices are summed, as in standard notation. For the convention of this section, we suppress the overall factor of $c^3/(16\pi G)$ in $S_\mathrm{PF}$, keeping the measure and source factors. To work in the Hamiltonian formulation, we separate temporal components from spatial ones. After integrating by parts, the Pauli-Fierz Lagrangian (up to total derivatives) is recast into the quadratic expression
\begin{equation}
    \begin{aligned}
        \mathcal{L}^\mathrm{L}&=\frac{1}{4c^2}\left[\left(\dot h^\mathrm{L}_{ij}-2\partial_{(i}h^\mathrm{L}_{j)0}\right)^2-\left(\dot h^\mathrm{L}_{ii}-2\partial_i h^\mathrm{L}_{i0}\right)^2\right]-\frac{1}{2}h^\mathrm{L}_{ij}R^\mathrm{L,lin}_{ij}\\&\quad+\frac{1}{4}\left(h^\mathrm{L}_{ii}-\frac{2h^\mathrm{L}_{00}}{c^2}\right)R^\mathrm{L,lin}_{jj}+\frac{8\pi G}{c^4}\left(h^\mathrm{L}_{ij}T^\mathrm{L}_{ij}-\frac{2}{c^2}h^\mathrm{L}_{0i}T^\mathrm{L}_{0i}+\frac{1}{c^4}h^\mathrm{L}_{00}T^\mathrm{L}_{00}\right) \,,
    \end{aligned}
\end{equation}
where we defined
\begin{equation}\label{defs3D}
    \begin{aligned}
        2R^{\mathrm{L},\mathrm{lin}}_{ij}\coloneqq2\partial_{(i}\partial^kh^\mathrm{L}_{j)k}-\Delta h^\mathrm{L}_{ij}-\partial_i\partial_j h^\mathrm{L}_{kk}\;, \quad R^{\mathrm{L},\mathrm{lin}}_{ii}=\partial_i\partial_j h^\mathrm{L}_{ij}-\Delta h^\mathrm{L}_{ii} \,,
    \end{aligned}
\end{equation}
and the parenthesis has weight $1/2$. The equations of motion are
\begin{equation}
    \begin{aligned}
        \partial_t\left(\dot h^\mathrm{L}_{ij}-\delta_{ij}\dot h^\mathrm{L}_{kk}-2\partial_{(i}h^\mathrm{L}_{j)0}+2\delta_{ij}\partial_kh^\mathrm{L}_{k0}\right)&=-2c^2R^\mathrm{L,lin}_{ij}+c^2\delta_{ij}R^\mathrm{L,lin}_{kk}\,, \\
        &\quad +(\delta_{ij}\Delta-\partial_{i}\partial_j)h^\mathrm{L}_{00} +\frac{16\pi G}{c^2}T^\mathrm{L}_{ij} \,, \\
        \partial_j\dot h^\mathrm{L}_{ij}-\partial_{i}\dot h^\mathrm{L}_{jj}-\Delta h^\mathrm{L}_{0i}+\partial_{i}\partial_jh^\mathrm{L}_{0j}&=\frac{16\pi G}{c^4}T^\mathrm{L}_{0i} \\
        R^\mathrm{L,lin}_{ii}&=\frac{16\pi G}{c^6}T^\mathrm{L}_{00} \,.
    \end{aligned}
\end{equation}
The canonical momentum conjugate to $h^\mathrm{L}_{ij}$ is
\begin{equation}
    \pi^\mathrm{L}_{ij}=\frac{\delta\mathcal{L}^\mathrm{L}}{\delta (\partial_0 h^\mathrm{L}_{ij})}=\frac{1}{2c^2}\left[\dot h^\mathrm{L}_{ij}-2\partial_{(i}h^\mathrm{L}_{j)0}-\delta_{ij}\left(\dot h^\mathrm{L}_{kk}-2\partial_kh^\mathrm{L}_{k0}\right)\right] \,.
\end{equation}
There are also two primary constraints $\pi_{00} = 0$ and $\pi_{0i} =0 $ which are associated with the gauge symmetry, namely the linearised diffeomorphisms, of the theory. It is also useful to express the velocities $\dot h^\mathrm{L}_{ij}$ in terms of the momenta  
\begin{equation}
    \dot h^\mathrm{L}_{ij}=2\partial_{(i}h^\mathrm{L}_{j)0}+c^2\left(2\pi^\mathrm{L}_{ij}-\delta_{ij}\pi^\mathrm{L}_{kk}\right)\;.
\end{equation}
The Hamiltonian action is
\begin{equation}
    S_H^\mathrm{L}=\int c\,d^4x\left(\pi^\mathrm{L}_{ij}\dot h^\mathrm{L}_{ij}-\mathcal{H}^\mathrm{L}_\mathrm{can}-h^\mathrm{L}_{00}\mathcal{C}-{h^\mathrm{L}_{0i}}\mathcal{C}_i\right) \,,
\end{equation}
with the canonical pair $(h^\mathrm{L}_{ij},\pi^\mathrm{L}_{ij})$. Up to spatial total derivatives, the canonical Hamiltonian is
\begin{equation}
    \mathcal{H}^\mathrm{L}_\mathrm{can}= c^2\left[(\pi_{ij}^\mathrm{L})^2-\frac{1}{2}\left({\pi_{ii}^\mathrm{L}}\right)^2\right]+\frac{1}{2}h_{ij}^\mathrm{L}R^{\mathrm{L},\mathrm{lin}}_{ij}-\frac{1}{4}h_{ii}^\mathrm{L}R^\mathrm{L,lin}_{jj}-\frac{8\pi G}{c^4}h_{ij}^\mathrm{L}T_{ij}^\mathrm{L} \,,
\end{equation}
and the first-class constraints\footnote{The first-class constraints $\mathcal C \approx 0$  and $\mathcal C_i \approx 0$ follow from the stability condition of the primary constraints, $\dot{\pi}_{00} = 0$ and $\dot{\pi}_{0i}= 0$ respectively} are
\begin{equation}
    \mathcal{C}=\frac{1}{2c^2}\left(R^\mathrm{L,lin}_{ii}-\frac{16\pi G}{c^6}T^\mathrm{L}_{00}\right)\;, \quad \mathcal{C}_i=-2\left(\partial_j\pi^\mathrm{L}_{ij}-\frac{8\pi G}{c^6}T^\mathrm{L}_{0i}\right)\,.
\end{equation}
We will discuss the gauge transformations generated by these constraints in the Carrollian theories after taking the $c\to 0$ limit. Note that the same action may also be obtained by perturbing the ADM action for gravity around lapse $\mathcal N = 1- h^\mathrm{L}_{00}/ c^2$ and shift $\mathcal N_i = h^\mathrm{L}_{0i}$.
\par
We now outline two cases obtained in the $c\to0$ limit of the action $S^\mathrm{L}_H = c S_{Carr}$ keeping $G$ fixed:

\begin{itemize}
 \item \textit{Magnetic case:} For the magnetic Carroll action, we map the Lorentzian fields as
    \begin{equation}
        h^\mathrm{L}_{00}=c^2h_{00}\;, \quad h^\mathrm{L}_{i\mu}=h_{i\mu}\;, \quad \pi_{ij}^\mathrm{L}=\pi_{ij}\;, 
            \end{equation}
            and the different components of the source $T_{\mu\nu}$ as
            \begin{equation}
           T_{0\mu}^\mathrm{L}=c^6T_{0\mu}\;, \quad T_{ij}^\mathrm{L}=c^4T_{ij}\;.
            \end{equation}
    \item \textit{Electric case:} In the electric case, the Lorentzian fields and momenta are rescaled inversely as follows
    \begin{equation}
        h^\mathrm{L}_{\mu\nu}=ch_{\mu\nu}\;, \quad \pi_{ij}^\mathrm{L}=\frac{\pi_{ij}}{c}\;,
    \end{equation}
    and the sources as
     \begin{equation}
            T_{00}^\mathrm{L}=c^7T_{00}\;,\quad  T_{0i}^\mathrm{L}=c^5T_{0i}\;, \quad T_{ij}^\mathrm{L}=c^3T_{ij}\;.
            \end{equation}
\end{itemize}
Both branches have a leading-order Hamiltonian action of the form
\begin{equation}
    S_H=\int dt\,d^3x\,\left(\pi_{ij}\dot h_{ij}-\mathcal{H}_{Carr}-h_{00}\mathcal{C}-h_{0i}\mathcal{C}_i\right) \,,
\end{equation}
where $\mathcal{H}_\mathrm{Carr}$ is the branch-dependent Hamiltonian density
\begin{equation}
    \mathcal{H}_\mathrm{Carr}=\begin{cases}
        (\pi_{ij})^2-\frac{1}{2}(\pi_{ii})^2-8\pi Gh_{ij}T_{ij} \quad &\mathrm{E} \\
        \frac{1}{2}h_{ij}R^{\mathrm{lin}}_{ij}
        -\frac{1}{4}h_{ii}R^{\mathrm{lin}}_{jj}-8\pi G h_{ij}T_{ij} \quad &\mathrm{M}
    \end{cases}\,,
\end{equation}
and in both cases the constraints are given by
\begin{equation}
    \mathcal{C}=\frac{1}{2}R^\mathrm{lin}_{ii}-8\pi GT_{00}\;, \quad \mathcal{C}_i=-2\left(\partial_j\pi_{ij}-8\pi GT_{0i}\right)\;.
\end{equation}
The Lorentzian conservation condition $\partial^\mu T_{\mu\nu}^\mathrm{L}=0$ here becomes
\begin{equation}
    \dot T_{0i}=\partial_j T_{ij}\;, \quad \dot T_{00}=\begin{cases}
        \partial_i T_{0i} \quad &\mathrm{E}\\
        0 &\mathrm{M}
    \end{cases}\,,\;
\end{equation}
which are the conditions that preserve the constraints when we first impose them.

The canonical gauge generator
\begin{equation}
    \quad \mathcal{G}[\zeta]=\int d^3x\,\left(2\zeta_0\mathcal{C}+\zeta_i\mathcal{C_i}\right)\,,
\end{equation}
generates the linearised diffeomorphisms through the Poisson bracket
\begin{equation}
     \{h_{ij}(x^k),\pi_{mn}(y^k)\}=\delta_{i(m}\delta_{n)j}\delta^{(3)}(x^k-y^k)\;,
\end{equation}
with the convention $\delta_\zeta X=\{X,\mathcal{G}\}$. The gauge transformations of the fields in both the branches are as follows
\begin{equation}
    \label{eq:gauge transf_gravity}
    \delta_\zeta h_{ij}=2\partial_{(i}\zeta_{j)}\;, \quad \delta_\zeta\pi_{ij}=\left(\delta_{ij}\Delta-\partial_i\partial_j\right)\zeta_0 \,,
\end{equation}
which agree with~\cite[(6.7)]{Henneaux:2021yzg}. In addition, from the requirement that the Hamiltonian action is invariant, the transformations of the multipliers are found to be
\begin{equation}
    \label{eq:gauge_transf_linearised_gravity_metric}
    \delta_\zeta h_{00}=2\dot\zeta_0\;, \quad \delta_{\zeta}h_{0i}=\begin{cases}
        \dot\zeta_i+\partial_i\zeta_0 \quad &\mathrm{E}\\
        \dot\zeta_i &\mathrm{M}
    \end{cases} \,,
\end{equation}
and the sources are gauge invariant, $\delta_\zeta T_{\mu\nu}=0$.

Under Carroll boosts, the components of fields and momenta in both the sectors transform differently. In the magnetic case, the Carroll boost transformation laws are
\begin{equation}
    \label{eq:magnetic_gravity_boost}
    \begin{aligned}
            \delta_b h_{ij}&=b_kx^k\dot h_{ij}+2b_{(i}h_{j)0} \,,\\ \delta_b\pi_{ij}&=b_kx^k\dot\pi_{ij}+\left(\delta_{ij}b_k\partial_k-b_{(i}\partial_{j)}\right)h_{00} \\
            &\quad+\frac{1}{2}b_k\partial_kh_{ij}-b_k\partial_{(i}h_{j)k}-\frac{1}{2}\delta_{ij}b_k\partial_kh_{ll}+\delta_{ij}b_k\partial_lh_{lk} \,,
    \end{aligned}
\end{equation}
whereas in the electric case, they are
\begin{equation}
\label{eq:electric_gravity_boost}
    \begin{aligned}
         \delta_b h_{ij}&=b_kx^k\dot h_{ij}+2b_{(i}h_{j)0} \,,\\
        \delta_b\pi_{ij}&=b_kx^k\dot\pi_{ij}+\left(\delta_{ij}b_k\partial_k-b_{(i}\partial_{j)}\right)h_{00}\,.
    \end{aligned}
\end{equation}
The transformations of the Lagrange multipliers $h_{00}$ and $h_{0i}$ follow from the invariance of the actions under Carroll boosts. Interestingly, unlike in the case of scalars and gauge theory, the boost transformations of $h_{00}$ and $h_{0i}$ also differ between the electric and magnetic branches.
\begin{equation}
 \delta_{b}h_{00}=\begin{cases}
        b_i x^i\dot h_{00} \quad &\mathrm{E}\\
        b_i x^i\dot h_{00}+2b_ih_{0i} &\mathrm{M}
        \end{cases}\,, \quad 
   \delta_{b}h_{0i}=\begin{cases}
        b_j x^j\dot h_{0i}+b_ih_{00} \quad &\mathrm{E}\\
        b_j x^j\dot h_{0i} &\mathrm{M}
    \end{cases}\,.
\end{equation}
 Since we shall refer to these sections in our construction of the convolutional double copy, all relevant expressions for the scalar theory, linearised Yang-Mills and linearised gravity are summarized in Table~\ref{tab:EM_branches}.

\begin{table}[!htbp]
\centering
\begingroup
\fontsize{9.5pt}{11pt}\selectfont

\definecolor{DCblue}{HTML}{2A6FDB}
\colorlet{DCink}{DCblue!75!black}
\colorlet{DCfill}{DCblue!7}

\setlength{\tabcolsep}{5pt}
\setlength{\aboverulesep}{0pt}
\setlength{\belowrulesep}{0pt}
\setlength{\abovecaptionskip}{6pt}
\renewcommand{\arraystretch}{1.15}

\def\DCband#1{%
  \rowcolor{DCfill}
  \multicolumn{2}{c}{%
    \rule[-5pt]{0pt}{17pt}%
    {\color{DCink}\scshape #1}}\\}

\def\DCfirstsub#1{%
  \addlinespace[2pt]
  {\color{DCink}\itshape #1}
  &
  {\color{DCink}\itshape #1}\\
  \addlinespace[3pt]}

\def\DCsub#1{%
  \addlinespace[4pt]
  {\color{DCink}\itshape #1}
  &
  {\color{DCink}\itshape #1}\\
  \addlinespace[3pt]}

\def\DCmath#1{$\displaystyle #1$}

\begin{tabularx}{\linewidth}{
  >{\centering\arraybackslash}X
  >{\centering\arraybackslash}X}

\arrayrulecolor{black}\toprule[.35pt]
\addlinespace[5pt]
\textsc{Electric} & \textsc{Magnetic} \\
\addlinespace[5pt]


\DCband{Scalar theory}

\DCfirstsub{Rescalings}

\DCmath{(\phi^\mathrm{L},\pi^\mathrm{L},j^\mathrm{L})=(c\phi,c^{-1}\pi,c^{-1}j)}
&
\DCmath{(\phi^\mathrm{L},\pi^\mathrm{L},j^\mathrm{L})=(\phi,\pi,j)}
\\

\DCsub{Hamiltonian}

\DCmath{
  \mathcal H^\mathrm{E}_\mathrm{Carr}
  =\frac12\pi^2+j\phi}
&
\DCmath{
  \mathcal H^\mathrm{M}_\mathrm{Carr}
  =\frac12(\partial_i\phi)^2+j\phi}
\\

\DCsub{EoM}

\DCmath{\dot\phi=\pi,\quad \dot\pi=-j}
&
\DCmath{\dot\phi=0,\quad \dot\pi=\partial_i^2\phi-j}
\\

\addlinespace[10pt]


\DCband{Linearised Yang-Mills}

\DCfirstsub{Rescalings}

\DCmath{(A_\mu^\mathrm{L},\pi_i^\mathrm{L},j_0^\mathrm{L},j_i^\mathrm{L})=(cA_{\mu},c^{-1}\pi_{i},cj_{0},c^{-1}j_{i})}
&
\DCmath{(A_\mu^\mathrm{L},\pi_i^\mathrm{L},j_0^\mathrm{L},j_i^\mathrm{L})=(A_{\mu},\pi_{i},c^{2}j_{0},j_{i})}
\\

\DCsub{Hamiltonian}

\DCmath{
  \mathcal H^\mathrm{E}_\mathrm{Carr}
  =\frac12\pi_i\pi_i+j_iA_i}
&
\DCmath{
  \mathcal H^\mathrm{M}_\mathrm{Carr}
  =\frac14F_{ij}F_{ij}+j_iA_i}
\\

\DCsub{EoM}

\DCmath{
  \dot A_i-\partial_iA_0=\pi_i,
  \quad \dot\pi_i=-j_i}
&
\DCmath{
  \dot A_i-\partial_iA_0=0,
  \quad \dot\pi_i=\partial_jF_{ji}-j_i}
\\[3pt]

\DCmath{\partial_i\pi_i=-j_0}
&
\DCmath{\partial_i\pi_i=-j_0}
\\

\addlinespace[10pt]


\DCband{Linearised gravity}

\DCfirstsub{Rescalings}

\DCmath{
  (h_{\mu\nu}^\mathrm{L},\pi_{ij}^\mathrm{L})
  =(ch_{\mu\nu},c^{-1}\pi_{ij})}
&
\DCmath{
  (h_{00}^\mathrm{L},h_{i\mu}^\mathrm{L},\pi_{ij}^\mathrm{L})
  =(c^2h_{00},h_{i\mu},\pi_{ij})}
\\[3pt]

\DCmath{
  (T_{00}^\mathrm{L},T_{0i}^\mathrm{L},T_{ij}^\mathrm{L})
  =(c^7T_{00},c^5T_{0i},c^3T_{ij})}
&
\DCmath{
  (T_{0\mu}^\mathrm{L},T_{ij}^\mathrm{L})
  =(c^6T_{0\mu},c^4T_{ij})}
\\

\DCsub{Hamiltonian}

\DCmath{
  \mathcal H^\mathrm{E}_\mathrm{Carr}
  =(\pi_{ij})^2-\frac12({\pi_{ii}})^2
   -8\pi G h_{ij}T_{ij}}
&
\DCmath{
  \mathcal H^\mathrm{M}_\mathrm{Carr}
  =\frac12h_{ij}R^{\mathrm{lin}}_{ij}
   -\frac14h_{ii}R^{\mathrm{lin}}_{jj}
   -8\pi G h_{ij}T_{ij}}
\\

\DCsub{EoM}

\DCmath{\dot h_{ij}-2\partial_{(i}h_{j)0}=2\pi_{ij}-\delta_{ij}{\pi}_{kk},}
&
\DCmath{\dot h_{ij}-2\partial_{(i}h_{j)0}=0,}
\\[4pt]

\DCmath{
  \begin{aligned}[t]
    \dot\pi_{ij}
      &=\frac12(\delta_{ij}\Delta-\partial_i\partial_j)h_{00}\\[2pt]
      &\quad+8\pi G T_{ij}
  \end{aligned}}
&
\DCmath{
  \begin{aligned}[t]
    \dot\pi_{ij}
      &=-R^{\mathrm{lin}}_{ij}
        +\frac12\delta_{ij}R^{\mathrm{lin}}_{kk}\\[2pt]
      &\quad+\frac12(\delta_{ij}\Delta-\partial_i\partial_j)h_{00}
        +8\pi G T_{ij}
  \end{aligned}}
\\

\addlinespace[6pt]
\bottomrule[.35pt]
\end{tabularx}

\caption{Electric and magnetic Carroll contractions of scalar theory, linearised Yang-Mills and linearised gravity with sources. $R^\mathrm{lin}$ is defined in \eqref{defs3D}.}
\label{tab:EM_branches}

\endgroup
\end{table}

\section{Carroll double copy with scalars}
\label{sec:Carroll double copy with scalars}
In this section we construct double copy dictionaries directly between Carrollian theories, taking as a starting point a convolutional dictionary for Lorentzian theories~\cite{Anastasiou:2018rdx,Anastasiou:2014qba,Borsten:2020xbt,LopesCardoso:2018xes}. For a chosen electric or magnetic branch of the target theory, $c$-scaling compatibility gives constraints that select the allowed combinations of the parent branches and restrict the form of the dictionary itself. The latter is additionally constrained by the requirement of being Carroll covariant, and of mapping solutions of the parent equations of motion into solutions of the target ones.

We begin with discussing the double copy of two scalar theories, as it already shows all the subtleties of defining suitable convolutions satisfying the above requirements for the dictionary, and allows us to introduce the circle products that enter in the subsequent gauge and gravity constructions. We then apply this formalism to another simple example: the double copy pairing a scalar with a linearised YM field, before moving on to the double copy of two linearised YM fields in~\cref{sec:Carrollian double copy to linearised gravity}.

\subsection{Lorentzian double copy}

The Carrollian dictionaries constructed in this paper all have a common starting point, which is the respective convolution double copy in Minkowski spacetime. We introduce the definitions and fundamental properties we will use throughout, and then we specialise to formulating the double-copy of two scalar theories.

Let $f$ and $g$ be fields in the adjoint representation of independent gauge groups carrying colour indices $a$ and $\tilde b$, respectively. The convolutional double copy combines these fields by multiplying their Fourier modes at the same four-momentum. In position space, this corresponds to taking a spacetime convolution
\begin{equation}
    [f\ast_4g](x)\coloneqq\int d^4y\,f(x-y)g(y)\;.
\end{equation}
This expression still carries both colour indices, which can be contracted by introducing a biadjoint scalar $\varphi$, called the \textit{spectator}, via the ternary operation
\begin{equation}
    \label{def:circle_4}
    f\circ_4 g\coloneqq f^a\ast_4(\varphi^{-1})^{a\tilde b}\ast_4 g^{\tilde b}\;, \quad \varphi\ast_4\varphi^{-1}\ast_4\varphi=\varphi\;.
\end{equation}
We refer to this operation as the \textit{circle product}.\footnote{Throughout, we restrict to Lorentz fields for which these convolutions are well-defined and associative.} It takes as inputs the two external factors $f,g$ together with the spectator, which enters the product via its convolutional pseudo-inverse $\varphi^{-1}$ defined by the equation on the right. When the pseudo-inverse is not unique, its choice is part of the dictionary~\cite{Anastasiou:2014qba,Borsten:2019prq}. The spectator is necessary even when the parent fields do not have colour indices, as it ensures that the field produced by the circle product has the appropriate mass dimension and correct transformation rule under translations. Indeed, under a translation $a^\mu$ of the external factors and spectator, the inverse kernel transforms as $-a^\mu$, thus compensating the shift of one of the external factors. Moreover, one can prove that $\circ_4$ is Lorentz covariant. A useful property that we will use below is that the circle product satisfies the same derivative rule as the convolution,
\begin{equation}
    \label{eq:derivative_identity_circ_4}
    \partial_\mu(f\circ_4g)=(\partial_\mu f)\circ_4g=f\circ_4(\partial_\mu g) \,.
\end{equation}
Note that with $x^0=ct$, each integration measure of the two convolutions in the definition of the circle product gives a factor of $c$. For a given spectator $\varphi$, the corresponding $\varphi^{-1}$ gets a factor of $c^{-2}$, as the convolutional pseudo-inverse defining relation also contains two convolutions. Therefore, the powers of $c$ coming from the measure of the circle product cancel.

Here, we are interested in the convolutional double copy of two sourced scalar theories in the Hamiltonian formulation, with variables $(\phi^\mathrm{L},\pi^\mathrm{L},j^\mathrm{L})$ and $(\tilde\phi^\mathrm{L},\tilde\pi^\mathrm{L},\tilde j^\mathrm{L})$ respectively, where the superscript $\mathrm{L}$ indicates that we are in the Lorentzian regime and no Carroll limit has been taken yet. We consider a convolutional dictionary giving a third Lorentzian scalar theory  $(\Phi^\mathrm{L},\Pi^\mathrm{L},J^\mathrm{L})$. It reads\footnote{It is always possible to normalise the coefficient of the scalar dictionary to $1$.}
\begin{equation}
    \label{eq:Lorentzian_dictionary_scalar}
        \Phi^\mathrm{L}=\phi^\mathrm{L}\circ_4\tilde\phi^\mathrm{L}\;, \quad \Pi^\mathrm{L}=\alpha\pi^\mathrm{L}\circ_4\tilde\phi^\mathrm{L}+\beta\phi^\mathrm{L}\circ_4\tilde\pi^\mathrm{L}\;.
\end{equation}
In what follows, we use this dictionary off-shell, where the coefficients $\alpha,\beta$ are independent, and the momenta $\pi^\mathrm{L},\tilde\pi^\mathrm{L}$ are not replaced by the time derivatives of the corresponding fields. The reason to not consider the on-shell dictionary is that the equations of motion would constrain the coefficients $\alpha,\beta$ appearing in \eqref{eq:Lorentzian_dictionary_scalar}, thus restricting the momentum dictionary before testing its compatibility with the Carroll target theory. The source dictionary will be derived in a second step, by requiring that the resulting map respects the sourced equations of motion of the Carroll theories.

\subsection{\texorpdfstring{$c$}{c}-rescalings}
\label{sec:parent_rescalings}
At this point, we perform the Carroll contractions of the parent theories. The fact that the inputs of the convolutional double copy consist of two independent (Hamiltonian pairs of) fields, plus the spectator, allows us to choose their Carrollian branches independently. So, although the three factors share the same limit $c\to0$, the rescaling assigned to each parent field does not require the others to undergo the same contraction. This means that in principle we are not restricted to considering dictionaries involving two electric or two magnetic parents (and spectator), but we can investigate combinations involving mixed electric-magnetic Carrollian scalar theories, and we also have the additional freedom to choose the branch of the spectator entering the circle product. The only constraint is that the choices of parent branches must combine to give a definite limit for the target theory, and this restricts the possible combinations, as we show here.

We denote the Carrollian parent variables by $(\phi,\pi,j)$ and $(\tilde\phi,\tilde\pi,\tilde j)$, and the target ones by $(\Phi,\Pi,J)$. For each target branch, we require the field and momentum dictionaries to respect the corresponding rescaling
\begin{equation}
    \label{eq:target_rescalings_scalar}
    (\Phi^\mathrm{L},\Pi^\mathrm{L})=\begin{cases}
        (c\Phi,c^{-1}\Pi) \quad & \text{E} \\
        (\Phi,\Pi) \quad & \text{M} \\
    \end{cases} \,,
\end{equation}
where E and M denote the electric and magnetic limits, as in \cref{sec:EM_limits}. So, we substitute the parent and spectator rescalings into \eqref{eq:Lorentzian_dictionary_scalar} and we seek the combinations whose total $c$ factor is compatible with one of the two options in \eqref{eq:target_rescalings_scalar}. In particular, a combination is allowed only if both the expression for the field and its corresponding momentum match exactly the ones above, and if necessary we impose conditions on $\alpha,\beta$ to remove the terms in the momentum dictionary that are not compatible. If a combination of parent branches does not satisfy this requirement then we discard it. Note that \eqref{eq:Lorentzian_dictionary_scalar} is symmetric under the exchange of the two input theories, as the circle product $\circ_4$ is symmetric and the momentum dictionary simply relabels $\alpha\leftrightarrow\beta$. This means that it is sufficient to study the pairs E-E, E-M and M-M for the parent scalar theories, each with either spectator branch. The list of the resulting six branch combinations, together with the corresponding $c$ rescalings for the target theory, is collected in \cref{tab:scalar_offshell_contractions}, which also shows which of them are allowed or discarded based on the compatibility with \eqref{eq:target_rescalings_scalar}\footnote{Of course, the spectator can have its own conjugate momentum $\pi_\varphi$. We leave its incorporation into the double copy dictionary for future work.}.

\begin{table}[!htbp]
\centering
\begingroup
\small

\definecolor{DCblue}{HTML}{2A6FDB}
\colorlet{DCink}{DCblue!75!black}
\colorlet{DCfill}{DCblue!7}
\colorlet{DCline}{DCblue!25}

\setlength{\tabcolsep}{5pt}
\setlength{\aboverulesep}{0pt}
\setlength{\belowrulesep}{0pt}
\renewcommand{\arraystretch}{1.05}

\newcommand{\DCgroup}[1]{%
  {\color{DCink}\footnotesize\scshape #1}}
\newcommand{\DCgstrut}{\rule[-3pt]{0pt}{18pt}}
\newcommand{\DChstrut}{\rule[-4pt]{0pt}{21pt}}
\newcommand{\DCrstrut}{\rule[-4pt]{0pt}{19pt}}

\begin{tabular}[t]{ccc}
\arrayrulecolor{black}\toprule[.35pt]
\addlinespace[2pt]
\rowcolor{DCfill}
\multicolumn{3}{c}{\DCgstrut\DCgroup{Input branches}} \\
\DChstrut
$(\phi,\pi)$ & $\varphi$ & $(\tilde\phi,\tilde\pi)$ \\
\arrayrulecolor{DCline}\midrule[.4pt]
\addlinespace[2pt]
\DCrstrut $\boldsymbol{\mathrm E}$ & $\boldsymbol{\mathrm E}$ & $\boldsymbol{\mathrm E}$ \\
\DCrstrut $\boldsymbol{\mathrm E}$ & $\boldsymbol{\mathrm E}$ & $\boldsymbol{\mathrm M}$ \\
\DCrstrut $\mathrm M$ & $\mathrm E$ & $\mathrm M$ \\
\DCrstrut $\mathrm E$ & $\mathrm M$ & $\mathrm E$ \\
\DCrstrut $\boldsymbol{\mathrm E}$ & $\boldsymbol{\mathrm M}$ & $\boldsymbol{\mathrm M}$ \\
\DCrstrut $\boldsymbol{\mathrm M}$ & $\boldsymbol{\mathrm M}$ & $\boldsymbol{\mathrm M}$ \\
\addlinespace[2pt]
\arrayrulecolor{black}\bottomrule[.35pt]
\end{tabular}%
\hspace{7pt}%
\begingroup
\setlength{\tabcolsep}{9pt}
\begin{tabular}[t]{ccc}
\arrayrulecolor{black}\toprule[.35pt]
\addlinespace[2pt]
\rowcolor{DCfill}
\multicolumn{3}{c}{\DCgstrut\DCgroup{Rescaling factor for}} \\
\DChstrut
$\phi^\mathrm{L}\circ_4\tilde\phi^\mathrm{L}$
& $\alpha\,\pi^\mathrm{L}\circ_4\tilde\phi^\mathrm{L}$
& $\beta\,\phi^\mathrm{L}\circ_4\tilde\pi^\mathrm{L}$ \\
\arrayrulecolor{DCline}\midrule[.4pt]
\addlinespace[2pt]
\DCrstrut $c$      & $c^{-1}$ & $c^{-1}$ \\
\DCrstrut $1$      & $c^{-2}$ & $1$      \\
\DCrstrut $c^{-1}$ & $c^{-1}$ & $c^{-1}$ \\
\DCrstrut $c^2$    & $1$      & $1$      \\
\DCrstrut $c$      & $c^{-1}$ & $c$      \\
\DCrstrut $1$      & $1$      & $1$      \\
\addlinespace[2pt]
\arrayrulecolor{black}\bottomrule[.35pt]
\end{tabular}%
\endgroup%
\hspace{7pt}%
\begin{tabular}[t]{c}
\arrayrulecolor{black}\toprule[.35pt]
\addlinespace[2pt]
\rowcolor{DCfill}
\DCgstrut\DCgroup{Target branch} \\
\DChstrut $(\Phi,\Pi)$ \\
\arrayrulecolor{DCline}\midrule[.4pt]
\addlinespace[2pt]
\DCrstrut $\boldsymbol{\mathrm E}$ \\
\DCrstrut $\boldsymbol{\mathrm M},\ \alpha=0$ \\
\DCrstrut $\varnothing$ \\
\DCrstrut $\varnothing$ \\
\DCrstrut $\boldsymbol{\mathrm E},\ (\beta=0)^*$ \\
\DCrstrut $\boldsymbol{\mathrm M}$ \\
\addlinespace[2pt]
\arrayrulecolor{black}\bottomrule[.35pt]
\end{tabular}
\par
\endgroup

\caption{Rescalings of the off-shell scalar Lorentzian dictionary, induced by different combinations of parent and spectator branch choices. The first block shows the branches chosen for each input of the dictionary, while the central block contains the $c$ factors multiplying the target field $\Phi^\mathrm{L}$ and the two terms of its momentum $\Pi^\mathrm{L}$ given in \eqref{eq:Lorentzian_dictionary_scalar}. The final column contains the target branches compatible with \eqref{eq:target_rescalings_scalar}, together with the required conditions on $\alpha,\beta$. The symbol $\varnothing$ indicates that no target branch is admissible. The  star $*$ identifies a term that disappears in the $c\to0$ limit without requiring its Lorentzian coefficient to vanish. Rows obtained by exchanging the two copies are omitted.}
\label{tab:scalar_offshell_contractions}
\end{table}

Crucially, the spectator $\varphi^\mathrm{L}$ appears in the dictionary via its convolutional inverse, meaning that a factor of $c^n$ in its rescaling contributes $c^{-n}$ in the product $\circ_4$. This means not only that the spectator branch actively contributes to the double copy, but also that two fixed Carrollian limits of the input scalar theories can give different output limits depending on which spectator we choose, as shown in the table. We can summarise the role of $\varphi^\mathrm{L}$ in determining the target Carroll limit by placing the parent branches in the numerator and the spectator branch in the denominator\footnote{As they appear in momentum space, the fact that the spectator should be (pseudo-)inverted as a matrix does not affect the overall $c$-scaling.} of the following schematic fractions:
\begin{equation}
    \label{eq:schematic_branch_combinations_scalar}
    \frac{\text{E}\times\text{E}}{\text{E}},\ \frac{\text{E}\times\text{M}}{\text{M}}\ \to\ \text{E}\;, \qquad \frac{\text{E}\times\text{M}}{\text{E}},\ \frac{\text{M}\times\text{M}}{\text{M}}\ \to\ \text{M}\;, \qquad \frac{\text{E}\times\text{E}}{\text{M}},\ \frac{\text{M}\times\text{M}}{\text{E}}\ \to\ \varnothing\;.
\end{equation}
In each allowed combination, the spectator rescaling compensates the $c$ factors of a scalar field in the same limit. When neither of the parent field limits matches the spectator branch, the overall rescaling of the target theory is not compatible with \eqref{eq:target_rescalings_scalar} and therefore we discard that choice. This in particular means that the electric and magnetic limits of the target theory can each be obtained by two different combinations of parent Carroll limits. As mentioned above and displayed in the table, in the admissible cases with mixed parents we additionally have to set one of the two terms in $\Pi_c$ to zero, in order for the target branch to be well defined. Specifically, when the output is electric then the magnetic parent momentum is not allowed to appear in the dictionary, and vice versa. By contrast, when we take the same limit for the parent theories, $\alpha$ and $\beta$ remain unconstrained off-shell.

The resulting expressions are our provisional candidates for the double-copy dictionaries. To complete the construction, we must additionally require that they are well defined on the relevant on-shell Carrollian fields and that they are compatible with Carroll transformations. In particular, the EoM of the magnetic scalar imposes its time independence, which can spoil the ordinary spacetime convolution. We then need to adapt the definition of the circle products to accommodate the Carroll inputs, together with ensuring the Carroll covariance of the whole dictionary.

\subsection{Circle products and Carroll covariance}
\label{sec:circle_products}
The dictionaries discussed above are still defined in terms of the Lorentzian circle product $\circ_4$, whose definition requires an extension to be compatible with  static factors. In fact, the magnetic equation $\dot\phi=0$ (see \cref{tab:EM_branches}) implies that we must deal with time-independent fields entering the dictionary, although in general the momenta and sources remain time dependent even in the magnetic branch. To ensure that the circle product can accommodate static factors, we introduce a regularisation prescription and we then impose compatibility with Carroll transformations.

\subsubsection{Static factors}
To study the obstruction more effectively we go to Fourier space, in which the circle product \eqref{def:circle_4} reads
\begin{equation}
    \widehat{f\circ_4 g}=\hat f^a(\hat\varphi^{-1})^{a\tilde b}\hat g^{\tilde b}\;,
\end{equation}
where $f,g$ are the external factors and $\hat\varphi^{-1}=\widehat{\varphi^{-1}}$ denotes the multiplicative pseudo-inverse in Fourier space corresponding to the spectator's convolutional pseudo-inverse in position space. A factor $q(t,x^i)=q_3(x^i)$ independent of the Carrollian time coordinate, which we refer to as \textit{static}, has transform
\begin{equation}
    \label{eq:static_Fourier_profile}
    \hat q(k^0,k^i)=2\pi\delta(k^0)\hat q_3(k^i)\;.
\end{equation}
Thus, pairing Carrollian fields using $\circ_4$ produces a Fourier expression that can contain products of delta functions or, for a static spectator, their formal inverse. The problem is that these operations are not well-defined in distribution theory, which means that the formal products and possible cancellations in the circle product require a regularisation prescription. Note that we are not implying that a single static external factor automatically makes the circle product ill-defined: if the spectator is time-dependent, $\circ_4$ remains well-defined provided that the remaining Fourier kernel admits a definite restriction to $k^0=0$.

For a given factor $q(x)$ and a regulating profile $r(x)$, we define $q_\mathrm{reg}\coloneqq rq_3$ if $q=q_3$ is static, and $q_\mathrm{reg}\coloneqq q$ otherwise (there is no need to regularise factors that do depend on time, so we leave them unaffected). We use the same profile for every static factor, including the spectator, and initially we choose
\begin{equation}
    \label{def:time_regulator}
    r(x)=r_\epsilon(t)\coloneqq e^{-\epsilon|t|}\;, \quad \epsilon>0\;.
\end{equation}
The original factor is recovered by taking $\epsilon\to0^+$. Analogously, in Fourier space $\hat r_\epsilon(k^0)\to2\pi\delta(k^0)$,\footnote{Convergence is understood in the sense of tempered distributions.} reproducing \eqref{eq:static_Fourier_profile}. So, the effect of this regulator is to replace the delta function by a regular profile, at finite $\epsilon$. The spectator is regulated before taking its convolutional pseudo-inverse. Throughout, we use the symbol $\circ_\epsilon$ to denote the double convolution whose inputs are regulated with $r_\epsilon$
\begin{equation}
    \label{eq:time_reg_circle_product}
    f\circ_\epsilon g\coloneqq f^a_\mathrm{reg}\ast_4(\varphi^{-1}_\mathrm{reg})^{a\tilde b}\ast_4 g^{\tilde b}_\mathrm{reg}\;, \quad r=r_\epsilon \,,
\end{equation}
which is well defined at finite $\epsilon$. We combine the temporal regulator factors in this expression and only then we take the $\epsilon\to0^+$ limit. Whether this combined limit exists or not depends on the inputs of \eqref{eq:time_reg_circle_product}, as shown in \cref{tab:regulator_limit_cases}.
\begin{table}[!htbp]
\centering
\begingroup
\small

\definecolor{DCblue}{HTML}{2A6FDB}
\colorlet{DCink}{DCblue!75!black}
\colorlet{DCfill}{DCblue!7}
\colorlet{DCline}{DCblue!25}

\setlength{\tabcolsep}{7pt}
\setlength{\aboverulesep}{0pt}
\setlength{\belowrulesep}{0pt}
\renewcommand{\arraystretch}{1.05}

\newcommand{\DCgroup}[1]{%
  {\color{DCink}\footnotesize\scshape #1}}
\newcommand{\DCgstrut}{\rule[-3pt]{0pt}{18pt}}
\newcommand{\DChstrut}{\rule[-4pt]{0pt}{21pt}}
\newcommand{\DCrstrut}{\rule[-4pt]{0pt}{19pt}}

\begin{tabular}[t]{
  *{3}{>{\centering\arraybackslash}p{18pt}}
}
\arrayrulecolor{black}\toprule[.35pt]
\addlinespace[2pt]
\rowcolor{DCfill}
\multicolumn{3}{c}{\DCgstrut\DCgroup{Input factors}} \\
\DChstrut
$f$ & $\varphi$ & $g$ \\
\arrayrulecolor{DCline}\midrule[.4pt]
\addlinespace[2pt]
\DCrstrut $\mathrm T$ & $\mathrm T$ & $\mathrm T$ \\
\DCrstrut $\mathrm T$ & $\mathrm T$ & $\mathrm S$ \\
\DCrstrut $\mathrm T$ & $\mathrm S$ & $\mathrm T$ \\
\DCrstrut $\mathrm S$ & $\mathrm T$ & $\mathrm S$ \\
\DCrstrut $\mathrm T$ & $\mathrm S$ & $\mathrm S$ \\
\DCrstrut $\mathrm S$ & $\mathrm S$ & $\mathrm S$ \\
\addlinespace[2pt]
\arrayrulecolor{black}\bottomrule[.35pt]
\end{tabular}%
\hspace{10pt}%
\begingroup
\setlength{\tabcolsep}{10pt}
\begin{tabular}[t]{cc}
\arrayrulecolor{black}\toprule[.35pt]
\addlinespace[2pt]
\rowcolor{DCfill}
\multicolumn{2}{c}{\DCgstrut\DCgroup{Regularisation}} \\
\DChstrut
Obstruction & Regulator \\
\arrayrulecolor{DCline}\midrule[.4pt]
\addlinespace[2pt]
\DCrstrut --- & $1$ \\
\DCrstrut --- & $\hat r_\epsilon$ \\
\DCrstrut $1/\delta(k^0)$ & $\hat r_\epsilon^{-1}$ \\
\DCrstrut $\delta(k^0)^2$ & $\hat r_\epsilon^2$ \\
\DCrstrut $\delta(k^0)/\delta(k^0)$ & $1$ \\
\DCrstrut $\delta(k^0)^2/\delta(k^0)$ & $\hat r_\epsilon$ \\
\addlinespace[2pt]
\arrayrulecolor{black}\bottomrule[.35pt]
\end{tabular}%
\endgroup%
\hspace{10pt}%
\begin{tabular}[t]{cc}
\arrayrulecolor{black}\toprule[.35pt]
\addlinespace[2pt]
\rowcolor{DCfill}
\multicolumn{2}{c}{\DCgstrut\DCgroup{Regulated limit}} \\
\DChstrut
$\lim_{\epsilon\to0^+}f\circ_{\epsilon}g$ & Type \\
\arrayrulecolor{DCline}\midrule[.4pt]
\addlinespace[2pt]
\DCrstrut $f\circ_4 g$ & $\mathrm T$ \\
\DCrstrut $f\circ_4 g$ & $\mathrm S$ \\
\DCrstrut Divergent & --- \\
\DCrstrut Divergent & --- \\
\DCrstrut $f\circ_3 g$ & $\mathrm T$ \\
\DCrstrut $f\circ_3 g$ & $\mathrm S$ \\
\addlinespace[2pt]
\arrayrulecolor{black}\bottomrule[.35pt]
\end{tabular}
\par
\endgroup

\caption{Regulated limits of the circle product \eqref{eq:time_reg_circle_product} for time-dependent (T) and static (S) factors and spectator. The column ``Obstruction'' shows schematically the undefined operations involving $\delta(k^0)$ before we regularise, while the ``Regulator'' column shows the temporal factor remaining after cancellation at finite $\epsilon$. Configurations obtained by exchanging $f$ and $g$ are omitted.}
\label{tab:regulator_limit_cases}
\end{table}
To be precise, there are exactly two situations in which the regulated limit in general diverges: TST and STS (the letters indicate the time dependence of $f$, the spectator $\varphi$ and $g$, in this order). Their divergent nature follows, respectively, from
\begin{equation}
    \hat r_\epsilon(k^0)^{-1}=\frac{(k^0)^2+\epsilon^2}{2\epsilon}\;, \quad \int\frac{dk^0}{2\pi}\,\hat r_\epsilon(k^0)^2=\frac{1}{\epsilon}\;,
\end{equation}
where the first diverges at non-zero frequency, and the second gives a divergent weight at zero frequency. On-shell, electric scalar fields in general depend on time, while magnetic ones are static. Thus, the divergent limits correspond precisely to the combinations $\frac{\text{E}\times\text{E}}{\text{M}}
$ and $\frac{\text{M}\times\text{M}}{\text{E}}$ that we already excluded in \cref{tab:scalar_offshell_contractions}. The off-shell $c$-rescaling analysis in the previous section thus anticipates the input branch combinations that are forbidden if we want to evaluate the dictionary on Carroll solutions.

The remaining four combinations are compatible with both classifications. In these finite limits, the regulated product reduces to the ordinary circle products, namely $\circ_4$ or its three-dimensional counterpart $\circ_3$, which is defined by the same formula \eqref{def:circle_4} with $\ast_3$ in place of $\ast_4$, and the spectator inverse taken in the spatial directions. A sufficient condition for the existence of the regulated product in the TTS combination is that $f\ast_4\varphi^{-1}\in L^1(\mathbb{R}_t)$, with of course the spatial integrations also (absolutely) convergent. Since there is a static factor in the convolution, the time dependence is integrated out, and the result in this case is static. Finally, in the TSS case the regulator factors cancel against each other, so that $\circ_3$ is evaluated at fixed time. Its result then has the same time dependence as $f$, while in the SSS case all factors and the output are static.

Even though this regularisation was motivated by the magnetic Carroll contraction, it acts on the convolution itself and it does not intrinsically rely on the $c\to0$ limit. As such, it gives a prescription for extending the electric Carroll dictionaries and the original Lorentzian dictionary to time independent fields. An analogous procedure can be applied to other directions, to regulate independence of the corresponding coordinate.

\subsubsection{Carroll covariance}
The circle product used in our dictionary must also be compatible with Carroll spacetime transformations, in order for the resulting fields to be Carroll covariant. However, the profile $r_\epsilon(t)$ used above to regulate static fields produces an obstruction to the boost covariance of the product, in the specific combination of a time-dependent external scalar factor $f$ with static external scalar $g_3$ and spectator $\varphi_3$. To see this, compute the limit of the regulated product with such inputs:
\begin{equation}
    \label{eq:regulated_mixed_case_product}
    \lim_{\epsilon\to0^+}\left[f\circ_\epsilon g_3\right](t,x^i)=[f^a(t,\cdot)\ast_3K^a_3](x^i)\;, \quad K_3^a\coloneqq(\varphi_3^{-1})^{a\tilde b}\ast_3 g_3^{\tilde b} \,,
\end{equation}
where $\ast_3$ indicates an ordinary spatial convolution and $\varphi_3^{-1}$ is the corresponding convolutional pseudo-inverse. At fixed time, this expression is a standard three-dimensional convolution on slices $t=\mathrm{const}$, as the kernel $K_3$ is static too. For covariance, the transformed regulated product must be equal to the regulated product of the transformed inputs. However, under an infinitesimal boost \eqref{el_boost_scalar},
\begin{equation}
    \begin{aligned}
        \label{eq:boost_obstruction}
        &\left[\delta_b\lim_{\epsilon\to0^+}(f\circ_\epsilon g_3)-\lim_{\epsilon\to0^+}((\delta_bf)\circ_\epsilon g_3)\right](t,x^i)=b_j\int d^3y\,y^j\partial_tf^a(t,x^i-y^i)K^a_3(y^i)\;,
    \end{aligned}
\end{equation}
where we used the fact that $K_3$ does not transform.\footnote{Explicitly, the LHS of \eqref{eq:boost_obstruction} is $b_jx^j\partial_t\int d^3y\,f^a(t,x^i-y^i)K_3^a(y^i)-\int d^3y\,b_j(x^j-y^j)\partial_tf^a(t,x^i-y^i)K_3^a(y^i)$.} This expression is in general non-zero, spoiling the boost covariance. The issue is that boosting changes the constant time-slices on which the spatial convolution is taken. To solve this problem, we must allow the choice of time slicing in the convolution to transform as well. As we shall see, the resulting product simply amounts to taking an ordinary spatial convolution in a frame adapted to that slice. 

To this end, we choose a different regularisation profile:
\begin{equation}
    \label{def:regulator_r_e_v}
    r_{\epsilon,v}(t,x^i)\coloneqq e^{-\epsilon|t+v_jx^j|}\;, \quad \partial_\mu v_i=0\;,
\end{equation}
where the constant three-dimensional vector $v_i$ labels a family of profiles that contain $r_\epsilon(t)$ at $v_i=0$. We allow $v_i$ to transform under Carroll boosts, and we determine its transformation law by enforcing covariance of the regularisation prescription. When taking the $\epsilon\to0^+$ limit, the circle product regulated with the profile $r=r_{\epsilon,v}$ reduces to
\begin{equation}
    \label{def:circ_v}
    (f\circ_vg_3)(t,x^i)\coloneqq\int d^3y\,f^a(t+v_jy^j,x^i-y^i)K^a_3(y^i)\;,
\end{equation}
where $K_3$ is the static kernel defined above. In fact, the regulating profiles cancel, as before, but the dependence on $v_i$ survives in the time shift of $f$. As a consistency check, note that for $v_i=0$ we reproduce the spatial convolution at fixed time slices given in \eqref{eq:regulated_mixed_case_product}. Throughout, we will make use of the following useful identities:
\begin{equation}
    \label{eq:derivatives_identity_circ_v}
    (\partial_\mu f)\circ_v g_3=\partial_\mu(f\circ_vg_3)\;, \quad f\circ_v(\partial_ig_3)=D_i(f\circ_vg_3)\;,\quad D_i\coloneqq\partial_i-v_i\partial_t \,,
\end{equation}
so taking a spatial derivative on the static factor corresponds to taking a derivative $D_i$ of the product along the slice $t+v_ix^i=\mathrm{const}$. The product $\circ_v$ is boost covariant when
\begin{equation}
    \label{eq:covariance_condition}
    \delta_b(f\circ_vg_3)-(\delta_bf)\circ_vg_3-(\delta_bv_i)\frac{\partial}{\partial v_i}(f\circ_vg_3)=0\;,
\end{equation}
where the derivative with respect to $v_i$ is computed at fixed inputs. The first two terms are the difference between boosting the output and boosting $f$. An explicit computation gives
\begin{equation}
    \delta_b(f\circ_vg_3)-(\delta_bf)\circ_vg_3=b_i\frac{\partial}{\partial v_i}(f\circ_vg_3)\;,
\end{equation}
that matches \eqref{eq:boost_obstruction} at $v_i=0$, as expected. The third term in \eqref{eq:covariance_condition} accounts for the fact that $v_i$ itself transforms, and it is precisely the one that allows us to eliminate the obstruction above, if we require
\begin{equation}
    \label{eq:shift}
    \delta_bv_i=b_i\;.
\end{equation}
So, the product $\circ_v$ with $v_i$ satisfying this transformation is compatible with Carroll boosts.

As anticipated, such a product can be interpreted as an ordinary spatial convolution in a frame adapted to the chosen time slice. To see this explicitly, observe that $\circ_v$ can also be represented in terms of finite scalar boosts of the form \eqref{eq:scalar_carroll_boost},
\begin{equation}
    \label{eq:circ_v_finite_rep}
    f\circ_vg_3=\Lambda_v\left[(\Lambda_{-v}f)^a\ast_3K_3^a\right]\;,
\end{equation}
where $\ast_3$ is taken at fixed time. Here, $\Lambda_{-v}$ rewrites $f$ in adapted coordinates with time $\tau=t+v_ix^i$, so taking the spatial convolution at fixed $\tau$ means that we are sampling the original $f$ on a tilted slice. Then $\Lambda_v$ boosts the result back to the original coordinates. Thus, $v_i$ encodes the frame in which the equal-time convolution is taken. Again, by setting $v_i=0$ we recover integration at $t=\mathrm{const}$. Geometrically, the tangent planes of the tilted slice are the kernel of the one form $k_v\coloneqq d t+v_id x^i$, which also satisfies $k_v(\partial_t)=1$. This form then is a choice of Carrollian ruling, an Ehresmann connection that selects spatial horizontal spaces in the complement of the Carrollian time direction $\partial_t$. Different choices of $v_i$ correspond to different choices of ruling, while the underlying Carrollian time direction is not affected~\cite{Ciambelli:2019lap,Ciambelli:2025unn}. Similar ideas linking Carrollian rulings to dressing maps are addressed in~\cite{Cresto:2024mne}.

When the static spectator $\varphi_3$ is magnetic and sourceless, we can give an explicit realisation of the constant vector $v_i$ defining the regulated product. Let $\pi_\varphi$ be the canonical momentum conjugate to $\varphi_3$, then we can define $v_i$ as the ratio between the total spatial momentum and total charge of the spectator,
\begin{equation}
    v_i\coloneqq\frac{P_i^{(\varphi)}}{E^{(\varphi)}}\;, \quad E^{(\varphi)}\coloneqq\frac{1}{2}\int d^3x\,(\partial_i\varphi_3)^2\;, \quad P_i^{(\varphi)}\coloneqq\int d^3x\,\pi_\varphi\partial_i\varphi\;.
\end{equation}
Indeed, under suitable regularity assumptions, these charges satisfy
\begin{equation}
    \dot E^{(\varphi)}=\dot P_i^{(\varphi)}=0\;, \quad \delta_bE^{(\varphi)}=0\;, \quad \delta_bP_i^{(\varphi)}=b_iE^{(\varphi)}\;,
\end{equation}
ensuring that $v_i$ is constant in spacetime and transforms as in \eqref{eq:shift}, as required. In particular, the role of the spectator momentum is to provide the non-trivial boost transformation, which the field $\varphi_3$ cannot supply. This is not the unique realisation for $v_i$, as covariance does not uniquely determine it. Nonetheless, this construction gives the required properties for $v_i$ entirely in terms of the Hamiltonian data of the spectator.

The shift of $v_i$ restoring the required boost transformation of the dressed product $\circ_v$ echoes the Stueckelberg mechanism described in~\cite{Nagy:2024dme,Nagy:2024jua,Nagy:2025hip}, which allows us to realise divergent asymptotic symmetry transformations at null infinity. In particular, the analogy is between the transformation law of $v_i$ and the one of the Goldstone-like dressing fields introduced to describe Yang-Mills overleading large gauge transformations.

To conclude this analysis, we observe that the limiting products are all compatible with rotations and spacetime translations. In particular, $v_i$ transforms as a spatial covector under rotations, while it is invariant under translations. Moreover, the other combinations of inputs not discussed here are already Carroll covariant so no obstruction has to be removed in those cases. Finally, we remark that the dressing procedure does not change the regularity behaviour of the $\epsilon\to0^+$ limits discussed above, so the classification of the admissible input combinations for the regulated product is not affected.

\subsubsection{Regulated circle product}
To sum up, by assigning the profile $r_{\epsilon,v}$ to static fields we can construct a circle product that allows us to combine Carrollian fields without breaking covariance under Carroll spacetime transformations. This regulated product is defined as
\begin{equation}
    \label{def:circ_reg}
    f\circ_\mathrm{reg}g\coloneqq\lim_{\epsilon\to0^+}\left[f^a_\mathrm{reg}\ast_4(\varphi^{-1}_\mathrm{reg})^{a\tilde b}\ast_4 g^{\tilde b}_\mathrm{reg}\right]\;,
\end{equation}
whenever the limit exists. Both external factors and the spectator are regulated according to
\begin{equation}
    q_\mathrm{reg}(x)\coloneqq\begin{cases}
        e^{-\epsilon|t+v_jx^j|}q_3(x^i) \quad &\text{$q=q_3$ static,} \\
        q(x) \quad &\text{$q$ time dependent}
    \end{cases}
\end{equation}
with $\epsilon>0$ and $v_i$ constant such that $\delta_bv_i=b_i$ under an infinitesimal Carroll boost. The spectator is regulated before taking its convolutional pseudo-inverse, and all the factors in the RHS of \eqref{def:circ_reg} are combined and simplified before taking the limit. As shown in \cref{tab:summary_circ_product}, when it exists the regularised circle product reduces to $\circ_4$ when at least two of its inputs are time-dependent, and to $\circ_v$ as defined in \eqref{def:circ_v} when at least two inputs are static. Moreover, when all three inputs are static, then $\circ_v$ coincides with $\circ_3$ (whose definition is analogous to \eqref{def:circle_4}, with spatial three-dimensional convolutions $\ast_3$ in place of $\ast_4$).

\begin{table}[!htbp]
\centering
\begingroup
\small

\definecolor{DCblue}{HTML}{2A6FDB}
\colorlet{DCink}{DCblue!75!black}
\colorlet{DCfill}{DCblue!7}
\colorlet{DCline}{DCblue!25}

\setlength{\tabcolsep}{10pt}
\setlength{\aboverulesep}{0pt}
\setlength{\belowrulesep}{0pt}
\renewcommand{\arraystretch}{1.1}

\newcommand{\DChead}[1]{%
  {\color{DCink}\footnotesize\scshape #1}}
\newcommand{\DCrow}{\rule[-4pt]{0pt}{21pt}}

\begin{tabular}{ccccc}
\arrayrulecolor{black}\toprule[.35pt]
\addlinespace[2pt]

\rowcolor{DCfill}
\multicolumn{3}{c}{%
  \rule[-3pt]{0pt}{19pt}\DChead{Inputs}}
&
\multicolumn{2}{c}{\DChead{Output}}
\\

\rule[-4pt]{0pt}{23pt}
$f$ & $\varphi$ & $g$
& $f\circ_{\mathrm{reg}}g$ & Type
\\
\arrayrulecolor{DCline}
\cmidrule(lr){1-3}
\cmidrule(lr){4-5}
\addlinespace[2pt]

\DCrow $\mathrm T$ & $\mathrm T$ & $\mathrm T$
& $f\circ_4 g$ & $\mathrm T$ \\

\DCrow $\mathrm T$ & $\mathrm T$ & $\mathrm S$
& $f\circ_4 g$ & $\mathrm S$ \\

\DCrow $\mathrm T$ & $\mathrm S$ & $\mathrm S$
& $f\circ_v g$ & $\mathrm T$ \\

\DCrow $\mathrm S$ & $\mathrm S$ & $\mathrm S$
& $f\circ_v g$ & $\mathrm S$ \\

\addlinespace[2pt]
\arrayrulecolor{black}\bottomrule[.35pt]
\end{tabular}
\endgroup

\caption{Reduction of the regulated circle product \eqref{def:circ_reg} for all the possible combinations of its inputs for which it exists. The letters T and S stand for time-dependent and static, respectively. Combinations related by $f\leftrightarrow g$ are listed only once. In the last row, $\circ_v\equiv \circ_3$.}
\label{tab:summary_circ_product}
\end{table}

Thus, the products $\circ_4$, $\circ_v$ and $\circ_3$ all arise from the same regularisation prescription, which collapses to one of them based on the time dependence of its inputs. In particular, none of them depends on $\epsilon$, as the limit removes it, while $\circ_v$ is the only one that keeps its dependence on the choice of slice through $v_i$. Throughout the rest of the paper, all circle products in the dictionaries are understood to be $\circ_\mathrm{reg}$, but we will explicitly write either $\circ_4$ or $\circ_v$ based on each given combination of external factors and spectator.

\subsection{Scalar \texorpdfstring{$\times$}{x} scalar}
With the regulated circle product above, we now have all the ingredients to complete the candidate dictionaries between Carrollian scalar theories obtained from the $c$-rescaling analysis. We recall that our goal is to combine two Carrollian scalar theories $(\phi,\pi,j)$ and $(\tilde\phi,\tilde\pi,\tilde j)$ into a third theory $(\Phi,\Pi, J)$ of the same kind, using all the possible allowed combinations of the parent and spectator branches identified in \cref{sec:parent_rescalings}.

Building on the behaviour of $\circ_\mathrm{reg}$ when acting on different combinations of time-dependent and static factors, and using the EoM for the scalar field in the electric and magnetic branches, we conclude that the dictionary combining the two fields takes the form
\begin{equation}
    \Phi=\begin{cases}
        \phi\circ_4\tilde\phi \quad &\text{electric spectator,} \\
        \phi\circ_v\tilde\phi \quad &\text{magnetic spectator,}
    \end{cases}
\end{equation}
where the vector $v_i$ in the second line satisfies \eqref{eq:shift}. Indeed, choosing an electric spectator selects the allowed pairings where at least one of the parent theories is in the electric branch, meaning that at least two of the inputs are time-dependent, which in turn selects $\circ_4$ as the product for the dictionary. Similarly, a magnetic spectator ensures that at least two inputs are static, and selects $\circ_v$ instead. Similarly, one can obtain the dictionaries for the target momentum $\Pi$ in all the allowed combinations, which in some cases are additionally constrained by Carroll covariance, as we show below. Then the expressions to obtain the target source are completely determined by the Hamiltonian field dictionaries together with the equations of motion. On top of this, whenever the two copies belong to the same branch, we also require symmetry under the exchange of the two scalar copies.\footnote{This condition is automatically satisfied in the case of two electric scalars.} The resulting dictionaries are collected in \cref{tab:scalar_scalar_dictionaries}.

\begin{table}[!htbp]
\centering
\begingroup
\small

\definecolor{DCblue}{HTML}{2A6FDB}
\colorlet{DCink}{DCblue!75!black}
\colorlet{DCfill}{DCblue!7}

\setlength{\tabcolsep}{6pt}
\setlength{\aboverulesep}{0pt}
\setlength{\belowrulesep}{0pt}
\renewcommand{\arraystretch}{1.12}

\def\DCtablewidth{\linewidth}

\def\DChead#1{%
  \rule[-4pt]{0pt}{20pt}#1}

\def\DCmath#1{%
  \parbox[c][42pt][c]{\linewidth}{\centering $#1$}}

\def\DCmap#1#2#3#4{%
  \DCmath{\displaystyle
    \frac{\mathrm{#1}_{\phi}\times\mathrm{#2}_{\tilde\phi}}
         {\mathrm{#3}_{\varphi}}
    \to\mathrm{#4}_\Phi}}

\def\DCband#1{%
  \noalign{%
    \vskip 3pt
    \hbox to \DCtablewidth{%
      \color{DCfill}%
      \leaders\vrule height 2.4ex depth 1.1ex
        \hskip\DCtablewidth
      \hskip-\DCtablewidth
      \hskip 6pt
      \color{DCink}\footnotesize\scshape #1\hfil}%
    \vskip 3pt}}

\begin{tabularx}{\DCtablewidth}{
  >{\hsize=.9\hsize\linewidth=\hsize
    \centering\arraybackslash}X
  >{\hsize=.7\hsize\linewidth=\hsize
    \centering\arraybackslash}X
  >{\hsize=1.2\hsize\linewidth=\hsize
    \centering\arraybackslash}X
  >{\hsize=1.2\hsize\linewidth=\hsize
    \centering\arraybackslash}X
}

\arrayrulecolor{black}\toprule[0.7pt]
\DChead{\textsc{Map}}
& \DChead{$\Phi$}
& \DChead{$\Pi$}
& \DChead{$J$}
\\

\DCband{Electric target}

\DCmap{E}{E}{E}{E}
&
\DCmath{\phi\circ_4\tilde\phi}
&
\DCmath{\pi\circ_4\tilde\phi=\phi\circ_4\tilde\pi}
&
\DCmath{j\circ_4\tilde\phi=\phi\circ_4\tilde j}
\\

\DCmap{E}{M}{M}{E}
&
\DCmath{\phi\circ_v\tilde\phi}
&
\DCmath{\pi\circ_v\tilde\phi}
&
\DCmath{j\circ_v\tilde\phi}
\\

\DCband{Magnetic target}

\DCmap{E}{M}{E}{M}
&
\DCmath{\phi\circ_4\tilde\phi}
&
\DCmath{\phi\circ_4\tilde\pi}
&
\DCmath{\phi\circ_4\tilde j}
\\

\DCmap{M}{M}{M}{M}
&
\DCmath{\phi\circ_v\tilde\phi}
&
\DCmath{\tfrac12(\pi\circ_v\tilde\phi+\phi\circ_v\tilde\pi)}
&
\DCmath{\tfrac12(j\circ_v\tilde\phi+\phi\circ_v\tilde j)}
\\

\bottomrule[0.7pt]
\end{tabularx}
\endgroup

\caption{Scalar double-copy dictionaries. The first column contains the combinations of parent theories and spectator branches in the schematic notation introduced in \cref{sec:parent_rescalings}. The subscripts identify which input the branch refers to, and the branch after each arrow refers to the output scalar theory $(\Phi,\Pi,J)$. Equalities in the first row hold on shell. The $\mathrm{M}\times\mathrm{E}$ dictionaries follow by exchanging the copies.}
\label{tab:scalar_scalar_dictionaries}
\end{table}

To illustrate how the above requirements determine the dictionaries, we work out in detail the case in which the scalar copies and the spectator are all in the magnetic branch. Recalling the last row of \cref{tab:scalar_offshell_contractions}, the candidate dictionary is
\begin{equation}
    \Phi=\phi\circ_v\tilde\phi\;, \quad \Pi=\alpha\pi\circ_v\tilde\phi+\beta\phi\circ_v\tilde\pi\;.
\end{equation}
Here, both $\phi$ and $\tilde\phi$ are static, meaning that the product in $\Phi$ actually reduces to $\circ_3$. By contrast, their momenta and sources may depend on time, so the regulated product in the expression for $\Pi$ specialises to $\circ_v$. As we demonstrated, the circle products have the right Carroll covariance properties ensuring that the target scalar field $\Phi$ transforms correctly. However, in the magnetic branch the momentum does not transform as a scalar, rather its Carroll boost transformation \eqref{eq:Carroll_boost_magnetic_momentum} has an additional term containing the gradient of the field. To test compatibility of the dictionary with boost covariance, along the lines of \eqref{eq:covariance_condition} we compute
\begin{equation}
    \delta_b\Pi-\alpha(\delta_b\pi)\circ_v\tilde\phi-\beta\phi\circ_v(\delta_b\tilde\pi)-b_i\frac{\partial\Pi}{\partial v_i}=(1-\alpha-\beta)b^i\partial_i\Phi\;,
\end{equation}
using the time-independence of the magnetic fields to rearrange spatial gradients. Thus, magnetic momentum covariance under Carroll boosts forces the constraint $\alpha+\beta=1$. Symmetry under exchange of the two scalar copies, $\pi\circ_v\tilde\phi\leftrightarrow\phi\circ_v\tilde\pi$, means that we must additionally impose $\alpha=\beta$, which gives the coefficient $1/2$ appearing in the last row of \cref{tab:scalar_scalar_dictionaries}. The source dictionary then follows from the equations of motion. In particular, the first equation, $\dot\Phi=0$, is automatically satisfied as both scalar inputs are static, while the second gives
\begin{equation}
    \dot\Pi=\frac{1}{2}\left[(\partial_i^2\phi-j)\circ_v\tilde\phi+\phi\circ_v(\partial_i^2\tilde\phi-\tilde j)\right]=\partial_i^2\Phi-\frac{1}{2}\left(j\circ_v\tilde\phi+\phi\circ_v\tilde j\right)\;,
\end{equation}
where we used the momentum dictionary just derived. Comparing with the EoM for the target momentum, $\dot\Pi=\partial_i^2\Phi-J$, determines the source shown in the table, whose scalar boost transformation follows from the covariance of the regulated product.

In a similar way one can build all the other dictionaries in \cref{tab:scalar_scalar_dictionaries}, which map the complete Hamiltonian data (fields, momenta and sources) to solutions of the target theory, satisfying the required Carroll transformation laws. The magnetic case shows how checking the scalar field alone is not sufficient, as boost covariance of the momentum imposes additional constraints.

\begin{figure}[!htbp]
\centering
\begingroup

\definecolor{DCblue}{HTML}{2A6FDB}
\colorlet{DCink}{DCblue!75!black}

\begin{tikzpicture}[
  x=\linewidth,
  y=1cm,
  endpoint/.style={
    draw=DCblue!30,
    fill=DCblue!7,
    text=DCink,
    line width=.35pt,
    text width=.15\linewidth,
    minimum height=1cm,
    inner sep=3pt,
    outer sep=0pt,
    align=center,
    font=\small
  },
  condition/.style={
    text=black,
    text width=.145\linewidth,
    align=center,
    font=\footnotesize,
    inner sep=0pt,
    anchor=north
  },
  marker/.style={
    circle,
    fill=black!85,
    inner sep=0pt,
    minimum size=2.5pt
  },
  flow/.style={
    -{Stealth[length=1.6mm]},
    draw=black!55,
    line width=.4pt
  }
]

\node[endpoint] (start) at (.085,0)
  {Off-shell\\Lorentzian\\dictionary};

\node[endpoint] (finish) at (.915,0)
  {On-shell\\Carrollian\\dictionaries};

\draw[flow] (start.east) -- (finish.west);

\foreach \x in {.27,.42,.57,.72}{
  \node[marker] at (\x,0) {};
}

\node[condition] at (.27,-.17)
  {Carroll \\$c$-rescalings};

\node[condition] at (.42,-.17)
  {Product regularisation};

\node[condition] at (.57,-.17)
  {Dictionary\\covariance};

\node[condition] at (.72,-.17)
  {Target EoM \\ compatibility};

\end{tikzpicture}
\endgroup

\caption{Steps of the construction of Carrollian scalar dictionaries from an off-shell Lorentzian dictionary.}
\label{fig:carroll_dictionary_procedure}
\end{figure}
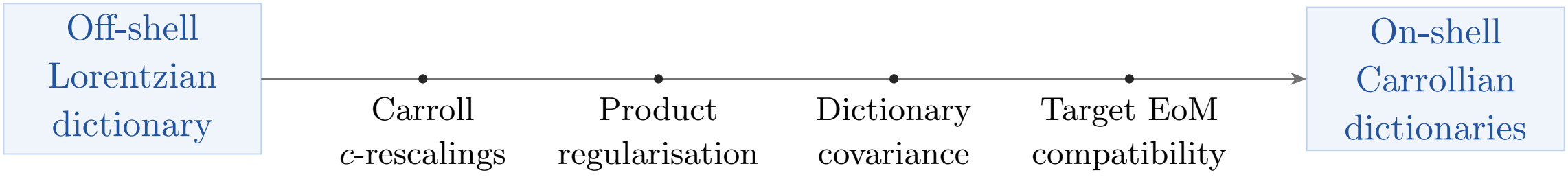

This concludes our treatment of the double copy between two scalars, for which we introduced all the machinery needed to address the more complicated cases involving linearised Yang-Mills and gravity. Before studying these cases, let us summarise the steps of the construction set up so far, illustrated in \cref{fig:carroll_dictionary_procedure}.
The starting point is the off-shell Lorentzian convolutional double copy, with a dictionary for both the target field and its Hamiltonian conjugate momentum. Using $c$-rescaling arguments, we identify all the branch combinations of the parent theories that give a well-defined Carroll limit of the target theory. Then, we introduce well-defined circle products arising from a common regularisation prescription for the four-dimensional product appearing in the Lorentzian dictionary. This step is necessary in order to use Carrollian fields as on-shell double-copy inputs. We then impose Carroll covariance of the dictionaries. Specifically, this requirement on scalar inputs can be implemented directly via dressing the temporal regulating profiles. Moreover, covariance on momentum dictionaries may further constrain their coefficients. Finally, we require compatibility with the equations of motion of the Carrollian target theory, which also allows us to determine the dictionaries for the sources without having to assume an independent ansatz for them. This concludes the procedure to obtain the desired Carrollian scalar double copy.

\subsection{Scalar \texorpdfstring{$\times$}{x} linearised YM}
We now apply the procedure developed above (summarised in \cref{fig:carroll_dictionary_procedure}) to the double copy of a Carrollian scalar theory $(\phi,\pi_\phi,j_\phi)$ and a linearised YM theory $(A_\mu,\pi_i,j_\mu)$. We denote the variables of the target Carrollian linearised YM theory with $(B_\mu,\Pi_i,J_\mu)$. In addition to the requirements imposed in the scalar example, here we also want the dictionaries to map gauge transformations of the input theories to gauge transformations of the output one.

As before, our starting point is the off-shell Lorentzian dictionary, which in this case reads\footnote{In principle, one could write a more general dictionary for $\Pi_i^\mathrm{L}$, including terms of the form $(\gamma/c^2)\dot\phi\circ_4 A_i+(\sigma/c^2)\phi\circ_4 F_{0i}$ with $\gamma,\sigma$ independent coefficients, motivated by the Lorentzian EoM. However, applying our construction to this more general ansatz gives the same surviving combinations \eqref{eq:surviving_combinations_scalar_YM} and the same Carroll dictionaries \eqref{eq:scalar_YM_dictionaries} derived starting from \eqref{eq:scalar_YM_dictionary_Lorentzial}.}
\begin{equation}
    \label{eq:scalar_YM_dictionary_Lorentzial}
    B^\mathrm{L}_\mu=\phi^\mathrm{L}\circ_4 A^\mathrm{L}_\mu\;, \quad \Pi^\mathrm{L}_i=\alpha\pi_\phi^\mathrm{L}\circ_4 A^\mathrm{L}_i+\beta\phi^\mathrm{L}\circ_4\pi^\mathrm{L}_i\;.
\end{equation}
As in the scalar case, we keep the coefficients $\alpha,\beta$ independent before taking the $c\to0$ contraction (imposing Lorentzian EoM would instead fix them). The sequential scaling, existence and consistency conditions imposed by our procedure are collected in \cref{tab:scalar_maxwell_consistency_steps}.
\begin{table}[!htbp]
\centering
\begingroup
\small

\definecolor{DCblue}{HTML}{2A6FDB}
\colorlet{DCink}{DCblue!75!black}
\colorlet{DCfill}{DCblue!7}

\setlength{\tabcolsep}{3pt}
\setlength{\aboverulesep}{0pt}
\setlength{\belowrulesep}{0pt}
\renewcommand{\arraystretch}{1}
\renewcommand{\tabularxcolumn}[1]{m{#1}}

\def\DChead#1{%
  \parbox[c][38pt][c]{\linewidth}{%
    \centering\normalfont\footnotesize
    \color{DCink}#1}}

\def\DCrow{\rule[-5pt]{0pt}{20pt}}
\def\DCmap#1#2#3{%
  $(\mathrm{#1},\mathrm{#2},\mathrm{#3})$}
\def\DCyes{$\checkmark$}
\def\DCskip{\textemdash}

\begin{tabularx}{\linewidth}{
  @{}
  >{\hsize=.85\hsize\linewidth=\hsize
    \centering\arraybackslash}X
  >{\hsize=1.45\hsize\linewidth=\hsize
    \centering\arraybackslash}X
  >{\hsize=1.05\hsize\linewidth=\hsize
    \centering\arraybackslash}X
  >{\hsize=.95\hsize\linewidth=\hsize
    \centering\arraybackslash}X
  >{\hsize=.80\hsize\linewidth=\hsize
    \centering\arraybackslash}X
  >{\hsize=.90\hsize\linewidth=\hsize
    \centering\arraybackslash}X
  @{}
}

\arrayrulecolor{black}\toprule[.35pt]

\rowcolor{DCfill}
\DChead{Branches\\$(\phi,\varphi,A_\mu)$}
&
\DChead{Carroll\\$c$-rescalings}
&
\DChead{Product\\existence}
&
\DChead{Dictionary\\covariance}
&
\DChead{Gauge\\compatibility}
&
\DChead{Target EoM\\compatibility}
\\
\addlinespace[2pt]

\DCrow \DCmap{E}{E}{E}
& $\mathrm E$
& $\circ_4$
& $\alpha=0$
& \DCyes
& $\beta=1$
\\

\DCrow \DCmap{E}{E}{M}
& $\mathrm M,\ \alpha=0$
& $\circ_4$
& $\beta=1$
& \DCyes
& \DCyes
\\

\DCrow \DCmap{M}{E}{E}
& $\mathrm M,\ \beta=0$
& $\circ_4$
& $\varnothing$
& \DCskip
& \DCskip
\\

\DCrow \DCmap{M}{E}{M}
& $\varnothing$
& \DCskip
& \DCskip
& \DCskip
& \DCskip
\\

\DCrow \DCmap{E}{M}{E}
& $\varnothing$
& \DCskip
& \DCskip
& \DCskip
& \DCskip
\\

\DCrow \DCmap{E}{M}{M}
& $\mathrm E,\ (\beta=0)^*$
& $\circ_{\mathrm{reg}}\to\infty$
& \DCskip
& \DCskip
& \DCskip
\\

\DCrow \DCmap{M}{M}{E}
& $\mathrm E,\ (\alpha=0)^*$
& $\circ_3$
& $\circ_v$
& \DCyes
& $\beta=1$
\\

\DCrow \DCmap{M}{M}{M}
& $\mathrm M$
& $\circ_3,\ \alpha=0$
& $\circ_v,\ \beta=1$
& \DCyes
& \DCyes
\\

\addlinespace[2pt]
\arrayrulecolor{black}\bottomrule[.35pt]
\end{tabularx}
\endgroup

\caption{Steps of the construction of the scalar $\times$ linearised YM double copy dictionaries. Each row is a combination of branches for the input fields $(\phi,\varphi,A_\mu)$, shown in the first column. Each of the remaining columns is a condition imposed in sequence, from left to right, following the procedure illustrated in \cref{fig:carroll_dictionary_procedure}. A checkmark denotes no further coefficient restriction, $\varnothing$ stands for exclusion, and dashes indicate stages not reached. The stars identify terms that disappear in the $c\to0$ limit without requiring their Lorentzian coefficients to vanish. Covariance of $\circ_v$ includes $\delta_bv_i=b_i$. In the last row, existence requires $\alpha=0$ because $\pi_\phi\circ_{\mathrm{reg}}A_i$ generically diverges.}
\label{tab:scalar_maxwell_consistency_steps}
\end{table}

There are a few differences compared to the scalar case above. First, since the parent theories are now different, we must study all eight branch combinations separately, because they are not related by exchange symmetry any more. Second, in general the magnetic gauge field is time-dependent, meaning that the existence of each product has to be studied solely from the time dependence of its inputs and it is not automatically linked to the specific branch. This can be seen, for example, in the case with electric scalar and magnetic spectator and gauge field, where the regulated product diverges even though the target electric branch is well-defined. Note that compatibility with gauge transformations does not give additional conditions, on top of Carroll covariance.

Of the eight possible combinations in \cref{tab:scalar_maxwell_consistency_steps}, only four survive all the steps:
\begin{equation}
    \label{eq:surviving_combinations_scalar_YM}
    \frac{\text{E}_\phi\times\text{E}_A}{\text{E}_\varphi},\ \frac{\text{M}_\phi\times\text{E}_A}{\text{M}_\varphi}\ \to\ \text{E}_B\;, \qquad \frac{\text{E}_\phi\times\text{M}_A}{\text{E}_\varphi},\ \frac{\text{M}_\phi\times\text{M}_A}{\text{M}_\varphi}\ \to\ \text{M}_B\;,
\end{equation}
where we used the schematic fraction notation, with the subscript of each branch label indicating the corresponding field. The pattern is clear: the scalar and spectator always belong to the same branch, while the target Carroll limit is determined by the YM branch alone. The $c$-rescalings of the scalar inputs ($\phi$ and $\varphi$) always compensate, so that the $c\to0$ contraction of the gauge field is the only responsible for the branch of the target theory. The resulting dictionaries are all of the form
\begin{equation}
    \label{eq:scalar_YM_dictionaries}
    B_\mu=\phi\circ_\mathrm{reg}A_\mu\;, \quad \Pi_i=\phi\circ_\mathrm{reg}\pi_i\;, \quad J_\mu=\phi\circ_\mathrm{reg}j_\mu\;,
\end{equation}
where $\circ_\mathrm{reg}=\circ_4$ when the spectator is electric, and $\circ_\mathrm{reg}=\circ_v$ when it is magnetic. In all four dictionaries, the scalar parent theory enters through $\phi$ alone, while its momentum and source never appear. In other words, once the scalar and the spectator are fixed, the double copy acts on the YM variables by the map $X\mapsto\phi\circ_\mathrm{reg}X$. Since $\circ_4$ and $\circ_v$ satisfy, respectively, \eqref{eq:derivative_identity_circ_4} and \eqref{eq:derivatives_identity_circ_v}, this map commutes with spacetime derivatives acting on $X$. From this property we immediately conclude that the linearised YM equations are carried into the corresponding target equations. The same property also implies that a gauge shift of $A_\mu$ can be mapped to a gauge shift of $B_\mu$ with target gauge parameter $\Lambda=\phi\circ_\mathrm{reg}\lambda$.

As a final remark, we note that if one fixes temporal gauge $A_0^\mathrm{L}=0$ in the initial Lorentzian dictionary and repeats the whole construction (imposing compatibility with the residual gauge transformations $\dot\lambda=0$), then one would find the same resulting dictionaries in \eqref{eq:scalar_YM_dictionaries} with temporal gauge imposed on the target theory, $B_0=0$. In the fully magnetic case, $A_i$ is then static, so the circle product in $B_i=\phi_3\circ_v A_i$ reduces to $\circ_3$. In the following section, when we take two copies of linearised YM, we shall impose temporal gauge on both copies to simplify the initial Lorentzian dictionary.

\section{Carroll double copy to linearised gravity}
\label{sec:Carrollian double copy to linearised gravity}
In this section, we build convolutional double copy dictionaries putting together two Carroll linearised Yang-Mills theories with Hamiltonian variables $(A_\mu,\pi_i,j_\mu)$ and $(\tilde A_\mu,\tilde \pi_i,\tilde j_\mu)$, respectively, to obtain Carroll linearised gravity $(h_{\mu\nu}, \pi_{ij}, T_{\mu\nu})$. As above, we combine the electric and magnetic branches of the parent theories and we work out the ones that provide a well defined Carroll contraction of the target theory. We restrict our attention to the metric sector, while the dilaton and the antisymmetric two-form dictionaries that arise in the double copy are left for future work.

As we shall explain below, it is convenient to choose temporal gauge on both the YM copies, namely
\begin{equation}
    A_0=\tilde A_0=0\;.
\end{equation}
On top of that, we fix Coulomb gauge on-shell for YM gauge fields in the magnetic branch.\footnote{In dictionaries containing a magnetic YM factor, we consider only gauge transformations preserving the domain on which the two-sided inverse spatial Laplacian $\Delta^{-1}$ is well-defined. Dictionaries retaining more general gauge transformations are possible, but are not discussed here.} This is compatible with temporal gauge, since the magnetic potentials are static.

First, we identify the admissible branch combinations via the same $c$-rescaling analysis presented in the previous section. Then, we discuss the electric and magnetic linearised gravity separately, imposing gauge and boost compatibility and checking the gravitational EoM. Finally, we collect the resulting dictionaries in~\cref{tab:YM_YM_final_dictionaries} at the end of the section.

\subsection{\texorpdfstring{$c$}{c}-rescalings}
We start from the off-shell Lorentzian dictionary
\begin{equation}
    \label{eq:Lorentzian_dictionary_gravity_temporal_gauge}
    h_{ij}^\mathrm{L}=2A^\mathrm{L}_{(i}\circ_4\tilde A^\mathrm{L}_{j)}\;, \quad \Pi_{ij}^\mathrm{L}=\alpha\pi_{(i}^\mathrm{L}\circ_4\tilde A^\mathrm{L}_{j)}+\beta A^\mathrm{L}_{(i}\circ_4\tilde\pi_{j)}^\mathrm{L}-\delta_{ij}\left(\alpha'\pi_{k}^\mathrm{L}\circ_4\tilde A^\mathrm{L}_{k}+\beta' A^\mathrm{L}_{k}\circ_4\tilde\pi_{k}^\mathrm{L}\right) \,,
\end{equation}
where $\alpha,\alpha',\beta,\beta'$ are free $c$-independent coefficients, since we do not go on-shell in the Lorentzian regime. The other components of the metric vanish, $h^\mathrm{L}_{0\mu}=0$, as a consequence of the temporal choice for the gauge fields. We apply the Carroll rescalings for the YM copies and the spectator introduced in~\cref{sec:EM_limits}, and we compare the resulting powers of $c$  in the Lorentzian dictionary with the target gravitational scalings
\begin{equation}
    \label{eq:target_rescalings_gravity}
    (h_{ij}^\mathrm{L},\Pi^\mathrm{L}_{ij})=\begin{cases}
        (ch_{ij},c^{-1}\Pi_{ij}) \quad & \text{E} \\
        (h_{ij},\Pi_{ij}) \quad & \text{M} \\
    \end{cases}\,.
\end{equation}
The results are collected in~\cref{tab:YM_YM_offshell_contractions}. The pattern mirrors that of the double copy between two scalars, since the canonical pairs $(A_i,\pi_i)$ and $(h_{ij},\Pi_{ij})$ have the same Carroll weights as the scalar ones, so the four surviving branch combinations are the same as in \eqref{eq:schematic_branch_combinations_scalar}. Labelling the branches in the order $(A,\varphi,\tilde A)$, the MEM and EME cases are again not allowed because the rescaling of $h^\mathrm{L}_{ij}$ gives a divergent or vanishing spatial metric in either target branch. By contrast, the EEM combination is allowed once the momentum dictionary is restricted by the condition $\alpha,\alpha'=0$, which removes the divergent terms in the vanishing $c$ limit. We remark that these conditions are obtained off-shell, while additional ones will be imposed to make the dictionaries compatible with symmetries and Carroll dynamics.

\begin{table}[!htbp]
\centering
\begingroup
\small

\definecolor{DCblue}{HTML}{2A6FDB}
\colorlet{DCink}{DCblue!75!black}
\colorlet{DCfill}{DCblue!7}
\colorlet{DCline}{DCblue!25}

\setlength{\tabcolsep}{5pt}
\setlength{\aboverulesep}{0pt}
\setlength{\belowrulesep}{0pt}
\renewcommand{\arraystretch}{1.05}

\newcommand{\DCgroup}[1]{%
  {\color{DCink}\footnotesize\scshape #1}}
\newcommand{\DCgstrut}{\rule[-3pt]{0pt}{18pt}}
\newcommand{\DChstrut}{\rule[-4pt]{0pt}{21pt}}
\newcommand{\DCrstrut}{\rule[-4pt]{0pt}{19pt}}

\begin{tabular}[t]{ccc}
\arrayrulecolor{black}\toprule[.35pt]
\addlinespace[2pt]
\rowcolor{DCfill}
\multicolumn{3}{c}{\DCgstrut\DCgroup{Input branches}} \\
\DChstrut
$(A_i,\pi_i)$ & $\varphi$ & $(\tilde A_i,\tilde\pi_i)$ \\
\arrayrulecolor{DCline}\midrule[.4pt]
\addlinespace[2pt]
\DCrstrut $\boldsymbol{\mathrm E}$ & $\boldsymbol{\mathrm E}$ & $\boldsymbol{\mathrm E}$ \\
\DCrstrut $\boldsymbol{\mathrm E}$ & $\boldsymbol{\mathrm E}$ & $\boldsymbol{\mathrm M}$ \\
\DCrstrut $\mathrm M$ & $\mathrm E$ & $\mathrm M$ \\
\DCrstrut $\mathrm E$ & $\mathrm M$ & $\mathrm E$ \\
\DCrstrut $\boldsymbol{\mathrm E}$ & $\boldsymbol{\mathrm M}$ & $\boldsymbol{\mathrm M}$ \\
\DCrstrut $\boldsymbol{\mathrm M}$ & $\boldsymbol{\mathrm M}$ & $\boldsymbol{\mathrm M}$ \\
\addlinespace[2pt]
\arrayrulecolor{black}\bottomrule[.35pt]
\end{tabular}%
\hspace{7pt}%
\begingroup
\setlength{\tabcolsep}{9pt}
\begin{tabular}[t]{ccccc}
\arrayrulecolor{black}\toprule[.35pt]
\addlinespace[2pt]
\rowcolor{DCfill}
\multicolumn{5}{c}{\DCgstrut\DCgroup{Rescaling factors}} \\
\DChstrut
$h^\mathrm{L}_{ij}$ & $\alpha$ & $\alpha'$ & $\beta$ & $\beta'$ \\
\arrayrulecolor{DCline}\midrule[.4pt]
\addlinespace[2pt]
\DCrstrut $c$      & $c^{-1}$ & $c^{-1}$ & $c^{-1}$ & $c^{-1}$ \\
\DCrstrut $1$      & $c^{-2}$ & $c^{-2}$ & $1$      & $1$      \\
\DCrstrut $c^{-1}$ & $c^{-1}$ & $c^{-1}$ & $c^{-1}$ & $c^{-1}$ \\
\DCrstrut $c^2$    & $1$      & $1$      & $1$      & $1$      \\
\DCrstrut $c$      & $c^{-1}$ & $c^{-1}$ & $c$      & $c$      \\
\DCrstrut $1$      & $1$      & $1$      & $1$      & $1$      \\
\addlinespace[2pt]
\arrayrulecolor{black}\bottomrule[.35pt]
\end{tabular}%
\endgroup%
\hspace{7pt}%
\begin{tabular}[t]{c}
\arrayrulecolor{black}\toprule[.35pt]
\addlinespace[2pt]
\rowcolor{DCfill}
\DCgstrut\DCgroup{Target branch} \\
\DChstrut $(h_{ij},\Pi_{ij})$ \\
\arrayrulecolor{DCline}\midrule[.4pt]
\addlinespace[2pt]
\DCrstrut $\boldsymbol{\mathrm E}$ \\
\DCrstrut $\boldsymbol{\mathrm M},\ \alpha=\alpha'=0$ \\
\DCrstrut $\varnothing$ \\
\DCrstrut $\varnothing$ \\
\DCrstrut $\boldsymbol{\mathrm E},\ (\beta=\beta'=0)^*$ \\
\DCrstrut $\boldsymbol{\mathrm M}$ \\
\addlinespace[2pt]
\arrayrulecolor{black}\bottomrule[.35pt]
\end{tabular}
\par
\endgroup

\caption{Rescalings of the off-shell linearised YM Lorentzian dictionary, with both YM fields in temporal gauge. The conventions are the same as \cref{tab:scalar_offshell_contractions}. The columns $\alpha,\alpha',\beta,\beta'$ give the rescaling factors of the corresponding terms in the momentum dictionary \eqref{eq:Lorentzian_dictionary_gravity_temporal_gauge}.}
\label{tab:YM_YM_offshell_contractions}
\end{table}

At this point, after rescaling and taking the $c\to0$ limit, we fix the Coulomb condition on the magnetic gauge fields on-shell. Since these fields are static, $\dot A_i=0$, we can impose
\begin{equation}
    \label{eq:coulomb_gauge}
    \partial_i A_i=0 \,,
\end{equation}
through a time-independent residual gauge transformation, that preserves temporal gauge.

As we showed in~\cref{sec:circle_products}, finiteness under the rescaling does not necessarily imply that the convolutional circle products are well defined when we apply them to solutions of the two Carroll parent theories. Our regularisation prescription allows us to obtain finite and well defined products, that can assume different expressions based on the time character of their inputs. Specifically, for an electric spectator we use $\circ_4$, while for a magnetic one the regulated product becomes $\circ_v$, which in turn reduces to $\circ_3$ when both external factors are static. In the following, we restrict to gauge transformations that preserve the domain of existence of our (regularised) convolution products, together with their corresponding derivational properties.

\subsection{Electric gravity}
The electric branch for linearised gravity is obtained from the combinations EEE and EMM, where the gauge choices above make it possible to express the dictionaries directly through the circle products introduced in the scalar case, as we shall see. The vanishing of $h_{0\mu}$ allows us to rearrange the electric equation of motion for the spatial metric (see~\cref{tab:EM_branches}) as
\begin{equation}
    \label{eq:electric_gravity_momentum_equation}
    \Pi_{ij}=\frac{1}{2}\left(\dot h_{ij}-\delta_{ij}\dot h_{kk}\right)\;,
\end{equation}
which fixes the remaining independent coefficients in the dictionary for the momentum \eqref{eq:Lorentzian_dictionary_gravity_temporal_gauge}, including normalisation and trace term. We now apply this relation to both the combinations giving electric gravity, and we check whether the resulting dictionary is compatible with symmetries.

\subsubsection{EEE}
Both the gauge fields and the spectator are time-dependent, so the regularised product is simply $\circ_4$. Using the derivative rule for the convolution, together with the Yang-Mills EoM and \eqref{eq:electric_gravity_momentum_equation}, we obtain the dictionary
\begin{equation}
    h_{ij}=2A_{(i}\circ_4\tilde A_{j)}\;, \quad \Pi_{ij}=\pi_{(i}\circ_4\tilde A_{j)}-\delta_{ij}\pi_k\circ_4\tilde A_k=A_{(i}\circ_4\tilde\pi_{j)}-\delta_{ij}A_k\circ_4\tilde\pi_k\;,
\end{equation}
Residual gauge transformations for the electric gauge fields induce the metric variation $\delta_\zeta h_{ij}=2\partial_{(i}\zeta_{j)}$ with
\begin{equation}
    \zeta_0=0\;, \quad \zeta_i=\lambda\circ_4\tilde A_i+A_i\circ_4\tilde \lambda\;.
\end{equation}
Since we are in temporal gauge, both $\dot\lambda$ and $\dot{\tilde\lambda}$ vanish, meaning that $\dot\zeta_i=0$ (preserving $h_{0\mu}=0$) and $\delta_\zeta\Pi_{ij}=0$. Moreover, temporal gauge simplifies the electric gravitational boost transformations \eqref{eq:electric_gravity_boost}, so that both $h_{ij}$ and $\Pi_{ij}$ behave as scalars under a Carroll boost, as the electric YM potentials do, see \eqref{eq:boost_electric_YM}. The boost covariance of $\circ_4$ is then sufficient to ensure boost compatibility of the dictionary. Finally, the source dictionary is determined from the remaining gravitational EoM, which give
\begin{equation}
    \begin{aligned}
        8\pi GT_{00}&=-\frac{1}{2}F_{ij}\circ_4\tilde F_{ij}\,,\\
        8\pi G T_{0i}&=-\frac{1}{2}\left(\pi_j\circ_4\tilde F_{ij}+F_{ij}\circ_4\tilde\pi_j\right)\,, \\
        8\pi GT_{ij}&=-j_{(i}\circ_4\tilde A_{j)}+\delta_{ij}j_k\circ_4\tilde A_k\;.
    \end{aligned}
\end{equation}

\subsubsection{EMM}
Here, the temporal gauge on the magnetic potential makes it static on-shell, as one can see from the magnetic EoM in~\cref{tab:EM_branches}, so the regulated circle product in this case becomes $\circ_v$, introduced in \eqref{def:circ_v}. As detailed in~\cref{sec:circle_products}, $v_i$ is a spacetime-constant vector transforming as $\delta_b v_i=b_i$ under infinitesimal Carroll boosts with parameter $b_i$. The terms with coefficients $\beta,\beta'$ disappear in the $c\to0$ limit, and \eqref{eq:electric_gravity_momentum_equation} additionally imposes $\alpha=\alpha'=1$, so the dictionary reads
\begin{equation}
    h_{ij}=2A_{(i}\circ_v\tilde A_{j)}\;, \quad \Pi_{ij}=\pi_{(i}\circ_v\tilde A_{j)}-\delta_{ij}\pi_k\circ_v\tilde A_k\;.
\end{equation}
Using the Coulomb gauge \eqref{eq:coulomb_gauge} on the magnetic leg, the residual YM electric gauge freedom maps to
\begin{equation}
    \zeta_0=0\;, \quad \zeta_i=\lambda\circ_v\tilde A_i\;.
\end{equation}
In the expression for $\zeta_i$, the circle product actually reduces to $\circ_3$, as both the external factors and the spectator are static, which makes $\zeta_i$ time-independent and $\delta \Pi_{ij}=0$, so again the electric gravitational gauge transformation is guaranteed, and it also preserves $h_{0\mu}=0$. The time dependence of the gravity fields is inherited from the electric gauge potential. 

Just as in the previous case, $h_{ij},\Pi_{ij},\pi_i,A_i$ transform like scalars under boosts, while here $\tilde A_i$ and the spectator are boost invariant. Therefore, the product $\circ_v$ is sufficient to reproduce the gravity boost transformations \eqref{eq:electric_gravity_boost} in temporal gauge. The electric gravity equations of motion fix the source dictionary,
\begin{equation}
    \begin{aligned}
        8\pi GT_{00}&=-\Delta\left(A_i\circ_v\tilde A_i\right)-v_ij_0\circ_v\tilde A_i\,,\\
        8\pi G T_{0i}&=-\partial_i\left(\pi_j\circ_v\tilde A_j\right)-\frac{1}{2}\left(j_0\circ_v\tilde A_i+v_jj_i\circ_v\tilde A_j\right)\,, \\
        8\pi GT_{ij}&=-j_{(i}\circ_v\tilde A_{j)}+\delta_{ij}j_k\circ_v\tilde A_k\;,
    \end{aligned}
\end{equation}
where $\Delta=\partial_i^2$ and the appearance of $v_i$ in the dictionary can be traced back to \eqref{eq:derivatives_identity_circ_v}. In general, it is possible to write more general metric, momentum and source dictionaries, if one removes the Coulomb gauge fixing on the magnetic factor. This requires a non-zero shift in the $h_{0i}$ component of the metric, together with the addition of suitable boost compensating terms.

\subsection{Magnetic gravity}
The branch combinations yielding magnetic linearised gravity are MMM and EEM. The first equation of motion for magnetic gravity in~\cref{tab:EM_branches} imposes $\dot h_{ij}=0$, in our setting in which $h_{0\mu}$ vanishes. As opposed to the electric case, in the magnetic branch the gravity momentum is no longer related to the time derivative of the metric, so the construction of its dictionary has to proceed differently. In particular, the momentum has a metric-dependent boost transformation
\begin{equation}
    \label{eq:momentum_boost_transformation_magnetic}
    \delta_b\Pi_{ij}=(b_kx^k)\dot\Pi_{ij}+B_b[h]_{ij}\;,
\end{equation}
which is equation \eqref{eq:magnetic_gravity_boost} in temporal gauge, where we introduced the notation
\begin{equation}
    B_b[h]_{ij}\coloneqq b_k\left(\frac{1}{2}\partial_kh_{ij}-\partial_{(i}h_{j)k}-\frac{1}{2}\delta_{ij}\partial_kh_{\ell\ell}+\delta_{ij}\partial_\ell h_{\ell k}\right)\,,
\end{equation}
for later convenience. Our double-copy construction must reproduce this metric-dependent contribution from the transformation of the YM variables in its dictionary. We first focus on the MMM case, where we introduce a tensor prescription for the product that allows us to write a dictionary compatible with the above boosts, and that will also be used in the EEM combination.

\subsubsection{MMM}
Both Yang-Mills potentials are static on-shell, which means that the dictionary for the metric is simply
\begin{equation}
    h_{ij}=2A_{(i}\circ_3\tilde A_{j)}\,,
\end{equation}
which is static as required. Since the two copies here are in the same branch, imposing exchange symmetry partially fixes the independent coefficients in the Carroll-contracted momentum dictionary \eqref{eq:Lorentzian_dictionary_gravity_temporal_gauge}, requiring $\alpha=\beta$ and $\alpha'=\beta'$. Each term in the dictionary contains a time-dependent momentum and static gauge field and spectator, so the regulated product gives $\circ_v$. Although this product provides the correct boost transformation for scalars, the momentum built from it, which we denote with $\Pi^\circ$, does not satisfy the required gravitational variation. In fact, compared to \eqref{eq:momentum_boost_transformation_magnetic} we have the obstruction
\begin{equation}
    \delta_b\Pi_{ij}^\circ-(b_kx^k)\dot\Pi_{ij}^\circ-B_b[h]_{ij}=b_k\left[\left(\alpha-\frac{1}{2}\right)\partial_kh_{ij}+(1-\alpha)\partial_{(i}h_{j)k}-\left(\alpha'-\frac{1}{2}\right)\delta_{ij}\partial_k h_{\ell\ell}\right]\;.
\end{equation}
This is non-vanishing for every choice of $\alpha$, which means that we must modify the momentum dictionary. Choosing $\alpha=\alpha'=1/2$, the only remaining obstruction to remove is the term
\begin{equation}
    \label{eq:obstruction_MMM}
    \frac{1}{2}b_k\partial_{(i}h_{j)k}\;.
\end{equation}
In other words, the boost transformation for $\Pi^\circ$ fails to match the metric-dependent part of the gravitational boost just for a factor of $1/2$ in one of its terms. We note that this obstruction can be traced back to the field-strength term in the boost transformation for the magnetic YM momentum,
\begin{equation}
    \label{eq:momentum_transf_ref}
    \delta_b\pi_i=(b_jx^j)\dot\pi_i+b_jF_{ji}\;.
\end{equation}

To compute the required extension of the momentum dictionary, we first work at $v=0$ and we introduce the spatial operators
\begin{equation}
    \label{def:spatial_operators_P}
    P^\pm_{ij}\coloneqq\delta_{ij}\pm\partial_i\Delta^{-1}\partial_j\;.
\end{equation}
Their action on a vector $X$ is
\begin{equation}
    P^+X=X^\perp+2X^\parallel\;, \quad P^-X=X^\perp\;,
\end{equation}
where $(P^\pm X)_i=P^\pm_{ij}X_j$, and $X^\perp$ and $X^\parallel$ are, respectively, the transverse and longitudinal parts of $X$. Throughout, we use the shorthand $P^\pm X_i\coloneqq(P^\pm X)_i$ to lighten the notation. For a gauge field in Coulomb gauge, we can apply the operator $P^+$ to the problematic term in \eqref{eq:momentum_transf_ref}, to get
\begin{equation}
    P^+_{ij}(b_kF_{kj})=b_k\partial_kA_i-2\partial_i(b_kA_k)\;.
\end{equation}
So if we replace each YM momentum in the $\Pi^\circ$ dictionary by $P^+\pi$, we obtain the correct longitudinal coefficient required by $B_b[h]$. The projector $P^-$ will be essential in the EEM case, where it removes an obstruction in the gauge variation of the electric YM field.

At this point, we have fixed the boost contribution at $v=0$. To extend this correction to arbitrary $v$, we must dress the spatial operators with $v$, to make sure that their derivatives transform correctly under boosts. In fact, we shall dress $P^\pm$ regardless of whether it acts on static or time-dependent fields. To do so, we represent its action on a generic vector $X$ as a four-dimensional convolution and apply the dressing to its kernel, which is
\begin{equation}
    \mathscr{P}^\pm(t,x^i)\coloneqq \delta(t)\left[P^\pm\delta^{(3)}\right](x^i)\;, \quad P^\pm X_i=\mathscr{P}^\pm_{ij}\ast_4 X_j\;.
\end{equation}
Recall that $P^\pm$ is a spatial operator, i.e.\ it acts at fixed time, and this is why the above kernel contains a temporal delta, while $P^\pm$ acts on the spatial coordinates. We dress this kernel using the same tilted time coordinate appearing in the scalar regulator $r_{\epsilon,v}$ defined in \eqref{def:regulator_r_e_v}\footnote{The dressing \eqref{eq:dressing_P} can be expressed in terms of the scalar dressing profile $r_{\epsilon,v}$ via $\mathscr{P}^\pm_v=(2\epsilon)^{-1}(\epsilon^2-\partial^2_t)(r_{\epsilon,v}[P^\pm\delta^{(3)}])$, where the time derivatives are taken distributionally.}:
\begin{equation}
    \label{eq:dressing_P}
    \mathscr{P}^\pm_v(t,x^i)\coloneqq\mathscr{P}^\pm(t+v_jx^j,x^i)=\delta(t+v_jx^j)\left[P^\pm\delta^{(3)}\right](x^i)\;.
\end{equation}
Again, here $P^\pm$ acts on the spatial delta, and the identities hold in the distributional sense. Convolving with this dressed kernel gives
\begin{equation}
    \label{def:Pv}
    (\mathscr{P}^\pm_v)_{ij}\ast_4 X_j=P^\pm_vX_i\;, \quad (P^\pm_v)_{ij}\coloneqq \delta_{ij}\pm D_i(D_kD_k)^{-1}D_j\;,
\end{equation}
where $D_i=\partial_i-v_i\partial_t$ as defined in \eqref{eq:derivatives_identity_circ_v}. Thus, the effect of the dressing on these operators is to replace every spatial derivative $\partial_i$ in the definition of $P^\pm$ by derivatives tangent to the tilted slice $t+v_ix^i=\mathrm{const}$. At this point we can insert this dressed kernel in one of the external factors of the regulated circle product \eqref{def:circ_reg}, without affecting the result of its limit. The operator $P^+_v$ constructed in this way ensures that the obstruction \eqref{eq:obstruction_MMM} is removed at arbitrary values of $v$.

The resulting dictionary is then
\begin{equation}
    \Pi_{ij}=\frac{1}{2}\left[P^+_v\pi_{(i}\circ_v\tilde A_{j)}+A_{(i}\circ_vP^+_v\tilde\pi_{j)}-\delta_{ij}\left(\pi_k\circ_v\tilde A_k+\tilde \pi_k\circ_v A_k\right)\right] \,.
\end{equation}
As expected, the shift $\delta_bv_i=b_i$ makes sure that the dressed product gives the required term proportional to $\dot\Pi$, while the effect of the $P^+_v$ operator is to combine the field strength terms in the YM boost into the required term $B_b[h]$, so that with this dictionary we have the correct boost transformation
\begin{equation}
    \delta_b\Pi_{ij}=(b_kx^k)\dot\Pi_{ij}+B_b[h]_{ij}\;.
\end{equation}
Finally, the magnetic gravity EoM determine the dictionary for the sources to be
\begin{equation}
    \begin{aligned}
        8\pi GT_{00}&=-\Delta\left(A_i\circ_3\tilde A_i\right)\,,\\
        8\pi GT_{0i}&=\frac{1}{2}\partial_j\left(P^+_v\pi_{(i}\circ_v\tilde A_{j)}+A_{(i}\circ_vP^+_v\tilde\pi_{j)}\right)-\frac{1}{2}\partial_i\left(\pi_{k}\circ_v\tilde A_{k}+A_{k}\circ_v\tilde\pi_{k}\right)\,,\\
        8\pi GT_{ij}&=-\frac{1}{2}\left[P^+_vj_{(i}\circ_v\tilde A_{j)}+A_{(i}\circ_vP^+_v\tilde j_{j)}-\delta_{ij}\left(j_k\circ_v\tilde A_k+A_k\circ_v\tilde j_k\right)\right] \\
        &\quad-\partial_i\partial_j\left(A_k\circ_3\tilde A_k\right)\;.
    \end{aligned}
\end{equation}

\subsubsection{EEM}
In this last case, both the untilded YM field and the spectator are electric, meaning that the product in the metric dictionary is the standard four-dimensional double convolution,
\begin{equation}
    h_{ij}=2A_{(i}\circ_4\tilde A_{j)}\;.
\end{equation}
On its convergence domain, $\circ_4$ moves time derivatives between its inputs, so the fact that $\tilde A_i$ is static ensures $\dot h_{ij}=0$, as required from the magnetic gravity EoM in temporal gauge, cf.~\cref{tab:EM_branches}. The residual gauge transformations come from the electric YM factor, giving $ \delta_\zeta h_{ij}=2\partial_{(i}\zeta_{j)}$ with
\begin{equation}
    \zeta_i=\lambda\circ_4\tilde A_i\;, \quad \dot\zeta_i=0\;.
\end{equation}
However, the double-copied gravitational momentum is not compatible with the required gauge transformation. To see why, note that this transformation satisfies the condition
\begin{equation}
    \partial_j\delta_\zeta\Pi_{ij}=0\;,
\end{equation}
as one can see from \eqref{eq:gauge transf_gravity}. Let $\Pi^\circ$ be the momentum defined with the dictionary \eqref{eq:Lorentzian_dictionary_gravity_temporal_gauge} after the $c$-rescaling step of~\cref{tab:YM_YM_offshell_contractions}, which sets $\alpha=\alpha'=0$, and after the $c\to0$ limit. It satisfies
\begin{equation}
    \label{eq:EEM_gauge_obstruction}
    \partial_j\delta_\zeta\Pi_{ij}^\circ=\frac{\beta}{2}\Delta(\lambda\circ_4\tilde\pi_i)+\left(\frac{\beta}{2}-\beta'\right)\partial_i\partial_j(\lambda\circ_4\tilde\pi_j)\;,
\end{equation}
which cannot vanish for arbitrary fields unless $\beta=\beta'=0$. Since this obstruction comes from the longitudinal part of the electric gauge field, it can be projected out by acting on the electric leg with the operator $P^-$ introduced in \eqref{def:spatial_operators_P}, so that $P^-\partial_i\lambda=0$. 

At this point, we must dress $P^-$ so that the spatial derivatives in its definition have the correct transformation under boosts. This is done using the exact same procedure described in the MMM case, which provides the dressed projector $P^-_v$ in \eqref{def:Pv}. Since the electric $\lambda$ is static, $P^-_v\partial_i\lambda=P^-\partial_i\lambda=0$\footnote{In particular, note that for a static $\lambda$ we have $D_i(D_kD_k)^{-1}D_j\partial_j\lambda=D_i(D_kD_k)^{-1}D_jD_j\lambda=D_i\lambda=\partial_i\lambda$.} and the gauge obstruction \eqref{eq:EEM_gauge_obstruction} is removed at arbitrary $v$.

Let $\Pi^\mathrm{proj}$ denote the gravity momentum with the electric YM fields projected with $P^-_v$. As it turns out, gauge compatibility and the EoM leave the coefficients $\beta,\beta'$ unfixed. We choose $\beta=\beta'=1$ to simplify the dictionary for the sources.\footnote{Keeping $\beta,\beta'$ arbitrary would not eliminate the remaining boost obstruction \eqref{eq:EEM_second_boost_obstruction}.} Although $\Pi^\mathrm{proj}$ now has the required variation under gauge transformations, it fails to satisfy the correct transformation under boosts. Compared to \eqref{eq:momentum_boost_transformation_magnetic} we have the obstruction
\begin{equation}
    \label{eq:EEM_second_boost_obstruction}
    \delta_b\Pi_{ij}^\mathrm{proj}-(b_kx^k)\dot\Pi_{ij}^\mathrm{proj}-B_b[h]_{ij}\simeq b_kP^-_vA_k\circ_4\partial_{(i}\tilde A_{j)} \,,
\end{equation}
where we used $\simeq$ to indicate that the equality holds up to a gauge transformation of the magnetic gravitational momentum.\footnote{Explicitly, $\delta_\zeta\Pi_{ij}=(\delta_{ij}\Delta-\partial_i\partial_j)\zeta_0$, see \eqref{eq:gauge transf_gravity}, with $\dot\zeta_0=0$. One can further modify the momentum dictionary by adding a static gauge term to obtain exact boost covariance. The source dictionary would not be affected by this modification.} To eliminate this mismatch, we add to the momentum dictionary a term whose boost variation is equal and opposite to the obstruction. We observe that the product multiplying $b_k$ is boost-covariant (thanks to the dressed projector and the circle product) and static (as one can move a time derivative to the static $\tilde A$). This makes it boost invariant, so the shift $\delta_bv_i=b_i$ allows us to cancel the obstruction by subtracting the same expression where we replace $b_i$ with $v_i$. The final dictionary for the momentum is then
\begin{equation}
    \Pi_{ij}\simeq P^-_vA_{(i}\circ_4\tilde\pi_{j)}-\delta_{ij}P^-_vA_{k}\circ_4\tilde\pi_{k}-v_kP^-_vA_{k}\circ_4\partial_{(i}\tilde A_{j)}\;.
\end{equation}
The expression on the right-hand side is invariant under the residual YM gauge transformations, which is consistent with choosing $\zeta_0=0$ in the gravitational gauge dictionary. Finally, the remaining equation of motion of magnetic gravity gives the dictionary for the sources:
\begin{equation}
    \begin{aligned}
        8\pi GT_{00}&=-\Delta\left(A_i\circ_4\tilde A_i\right)\,,\\
        8\pi GT_{0i}&=-\partial_i\left(P^-_vA_j\circ_4\tilde\pi_j\right)-\frac{1}{2}\left(P^-_vA_i\circ_4\tilde j_0+v_jP^-_vA_j\circ_4\tilde j_i\right)\,,\\
        8\pi GT_{ij}&=-P^-_vA_{(i}\circ_4\tilde j_{j)}+\delta_{ij}P^-_vA_k\circ_4\tilde j_k-\partial_i\partial_j\left(A_k\circ_4\tilde A_k\right)\;.
    \end{aligned}
\end{equation}

\subsection{Summary of the dictionaries}
The linearised gravity metric and momentum dictionaries constructed above are collected in~\cref{tab:YM_YM_final_dictionaries}.

\begin{table}[H]
\centering
\begingroup
\small

\definecolor{DCblue}{HTML}{2A6FDB}
\colorlet{DCink}{DCblue!75!black}
\colorlet{DCfill}{DCblue!7}

\setlength{\tabcolsep}{10pt}
\setlength{\aboverulesep}{0pt}
\setlength{\belowrulesep}{0pt}
\renewcommand{\arraystretch}{1.12}

\def\DChead#1{\rule[-4pt]{0pt}{20pt}#1}
\def\DCrow{\rule[-18pt]{0pt}{42pt}}
\def\DCmath#1{$\displaystyle #1$}

\def\DCmap#1#2#3#4{%
  \DCrow
  $\displaystyle
    \frac{\mathrm{#1}_{A}\times\mathrm{#2}_{\tilde A}}
         {\mathrm{#3}_{\varphi}}
    \to\mathrm{#4}_{h}$}

\def\DCband#1{%
  \rowcolor{DCfill}
  \multicolumn{3}{l}{%
    \rule[-4pt]{0pt}{19pt}%
    \color{DCink}\footnotesize\scshape #1}\\}

\resizebox{\linewidth}{!}{%
\begin{tabular}{ccc}

\arrayrulecolor{black}\toprule[.7pt]
\DChead{\textsc{Map}}
& \DChead{$h_{ij}$}
& \DChead{$\Pi_{ij}$}
\\

\DCband{Electric target}

\DCmap{E}{E}{E}{E}
& \DCmath{2A_{(i}\circ_4\tilde A_{j)}}
& \DCmath{
  \pi_{(i}\circ_4\tilde A_{j)}
  -\delta_{ij}\pi_k\circ_4\tilde A_k
  =
  A_{(i}\circ_4\tilde\pi_{j)}
  -\delta_{ij}A_k\circ_4\tilde\pi_k}
\\

\DCmap{E}{M}{M}{E}
& \DCmath{2A_{(i}\circ_v\tilde A_{j)}}
& \DCmath{
  \pi_{(i}\circ_v\tilde A_{j)}
  -\delta_{ij}\pi_k\circ_v\tilde A_k}
\\

\DCband{Magnetic target}

\DCmap{E}{M}{E}{M}
& \DCmath{2A_{(i}\circ_4\tilde A_{j)}}
& \DCmath{
  P^-_vA_{(i}\circ_4\tilde\pi_{j)}-\delta_{ij}P^-_vA_{k}\circ_4\tilde\pi_{k}-v_kP^-_vA_{k}\circ_4\partial_{(i}\tilde A_{j)}}
\\

\DCmap{M}{M}{M}{M}
& \DCmath{2A_{(i}\circ_3\tilde A_{j)}}
& \DCmath{
  \tfrac12\left[P^+_v\pi_{(i}\circ_v\tilde A_{j)}+A_{(i}\circ_vP^+_v\tilde\pi_{j)}-\delta_{ij}\left(\pi_k\circ_v\tilde A_k+\tilde \pi_k\circ_v A_k\right)\right]}
\\

\bottomrule[.7pt]
\end{tabular}%
}
\endgroup

\caption{Double-copy dictionaries for Carroll linearised gravity, using the branch notation and gauge conditions stated above, with $h_{0\mu}=0$. The equality in the first row holds on-shell. The EEM momentum is given modulo the static gauge completion
discussed in the text. The $\mathrm{M}\times\mathrm{E}$ dictionaries follow by exchanging the copies.}
\label{tab:YM_YM_final_dictionaries}
\end{table}

\section{Conclusions}
\label{sec:Conclusions}

We have shown how to take the Carrollian limit of the convolutional double copy, and give explicit dictionaries that construct the fields, their symmetries and their equations of motion. The Carrollian limit of the output field, electric or magnetic, is directly determined by the choice of limit in the double copy factors. 
\par
This also gives us the opportunity to extend the convolutional dictionary in a few different directions. Firstly, we incorporate the Hamiltonian formulation, which is particularly interesting in the magnetic limit where the relation between the field and its conjugate momentum becomes more complicated. Secondly, we introduce a regularisation procedure that allows for a well-defined convolution in the case where the inputs are independent of one of the coordinates\footnote{Which previously involved cancelling $\delta$-functions in momentum space.}. Finally, we have seen how dressing fields can play a key role in preserving Carrollian boosts via the dictionary.  
\par
A natural next step is to extend the dictionary to the BRST formulation, which allows us to reproduce the full set of degrees of freedom arising from the double copy of two gauge fields. In particular, we can disentangle the graviton from the dilaton field\footnote{We have not treated the dilaton in this article.} in order to construct black hole solutions in GR. In this context, there is a recent treatment of Carrollian limits in the BRST formulation, which interestingly finds no interactions in the electric limit~\cite{Martin:2026xvl}. Black holes in Carrollian theories have been studied in~\cite{Perez:2021abf,Ecker:2023uwm,Ecker:2026noz,Hansen:2021fxi,Kolar:2025ebv}; it would be interesting to see if they can be constructed from Carrollian gauge fields.

On the other hand, there exist other regimes where the double copy is clearly manifest,  notably the self-dual sector of YM and gravity~\cite{Monteiro:2011pc,Chen:2019ywi,Chen:2021chy,Monteiro:2022nqt,Monteiro:2013rya,Elor:2020nqe,Farnsworth:2021wvs,Skvortsov:2022unu,Campiglia:2021srh,Ben-Shahar:2022ixa,CarrilloGonzalez:2024sto,Bonezzi:2023pox,Berman:2018hwd,Miller:2025wpq,Miller:2026fgp,Miller:2024oza,Armstrong-Williams:2022apo,Borsten:2023paw,Monteiro:2022xwq,Lipstein:2023pih,Nagy:2022xxs,Krasnov:2021cva,Monteiro:2022lwm,Brown:2023zxm,Chowdhury:2024dcy,LopesCardoso:2024ttc,LopesCardoso:2025ldi,Correa:2024mub,Diaz-Jaramillo:2025gxw,Chowdhury:2026jpt,Dunajski:2026dfe}. We plan to study the Carroll double copy in this regime. We expect this to naturally bring subleading orders in $c$: the self-duality condition relates electric and magnetic components that enter at different orders in the Carroll expansion, and the strict \(c\to0\) limit loses the relation between the two chiralities\footnote{See~\cite{Fecko:2022shq}, where it was observed that the Hodge dual becomes nilpotent in the strict Carrollian limit.}. Retaining subleading fields would allow us to track these self-duality relations through the double copy, thus also understanding how to incorporate corrections to the leading Carrollian dictionaries.

\par 
Carroll field theories can also be obtained from their Lorentzian counterparts through null reduction, where the Lorentzian action in Minkowski spacetime is projected onto a null hypersurface, yielding a Carroll action in one lower dimension~\cite{Duval:2014uoa}. This procedure is particularly simple to implement for field theories in the light-cone formulation~\cite{Majumdar:2025juk,Majumdar:2024rxg,Bagchi:2024epw}. The light-cone gauge has also proved effective for several structures in scattering amplitudes, such as the KLT relations and MHV Lagrangians~\cite{Gorsky:2005sf,Ananth:2007zy}. It is therefore worth investigating whether the Lorentzian double copy structure survives null reduction and leads to the Carroll double copy dictionaries found here.
\par
Complementary to the $c\to 0$ contraction of the Poincar\'e group, there exists the opposite limit $c \to \infty$, which leads to the Galilean regime. In recent years, there have been many studies of the Galilean limit of the Poincar\'e group, including the construction of Galilean field theories, non-relativistic strings, hydrodynamics and black holes~\cite{Goldin_2001,Bagchi:2015qcw,Mehra:2021sfx,Festuccia:2016caf,Hartong:2022lsy,Aviles:2026qjl}. It would be interesting to see whether the Lorentzian double copy structure also admits a $c\to \infty$ limit, in which one could formulate a Galilean double copy for non-relativistic gravity.

\section*{Acknowledgements}
We would like to thank Luca Ciambelli, Javier Peraza and Romain Ruzziconi for useful discussions. SN is supported by STFC consolidated grant
UKRI4226. G.P. is funded by STFC Doctoral Studentship 2023.  SM is grateful for the hospitality of Perimeter Institute where part of this work
was carried out. Research at Perimeter Institute is supported in part by the Government
of Canada through the Department of Innovation, Science and Economic Development and
by the Province of Ontario through the Ministry of Colleges and Universities. This work
was supported by a grant from the Simons Foundation (1034867, Dittrich).

\bibliography{Ref_all_r_SD}
\bibliographystyle{utphys}

\end{document}